\documentclass[prb,twocolumn,showpacs]{revtex4}
\usepackage{graphicx}
\usepackage{amsmath}
\begin{document}
\bibliographystyle{prsty} 
\title{\emph{Ab initio} study of the migration of intrinsic defects in 3C-SiC}

\author{Michel Bockstedte}
\email{bockstedte@physik.uni-erlangen.de} 
\author{Alexander Mattausch}
\author{Oleg Pankratov} 
\affiliation{Lehrstuhl f. Theoretische
    Festk\"orperphysik, Universit\"at Erlangen-N\"urnberg, Staudtstr. 7\,B2,
    D-91054 Erlangen, Germany} \date{\today}
\begin{abstract}
  The diffusion of intrinsic defects in 3C-SiC is studied using an \emph{ab
    initio} method based on density functional theory.  The vacancies are
  shown to migrate on their own sublattice. The carbon split-interstitials and
  the two relevant silicon interstitials, namely the tetrahedrally
  carbon-coordinated interstitial and the $\langle110\rangle$-oriented
  split-interstitial, are found to be by far more mobile than the vacancies.
  The metastability of the silicon vacancy, which transforms into a
  vacancy-antisite complex in \emph{p}-type and compensated material, kinetically
  suppresses its contribution to diffusion processes.  The role of
  interstitials and vacancies in the self-diffusion is analyzed. Consequences
  for the dopant diffusion are qualitatively discussed. Our analysis
  emphasizes the relevance of mechanisms based on silicon and carbon
  interstitials.
\end{abstract}
\maketitle
\section{Introduction}
\label{sec:intro}
Silicon carbide is an important wide band gap semiconductor which offers a
variety of applications for high power and high frequency devices. During
device processing, various intrinsic defects are introduced that affect the
electronic properties of the material.  The mobile intrinsic defects, namely
vacancies and interstitials, play a pivotal role in self diffusion and dopant
diffusion as well as in the annealing of ion-implanted material.

Substitutional impurities are \emph{per se} immobile. They need vacancies or
interstitials as vehicles for the migration.  The rapid diffusion in the
course of annealing can considerably affect the implanted dopant profiles. For
example, a transient enhanced dopant diffusion was observed in
boron\cite{laube:99,janson:00} and aluminum\cite{usov:99} implanted 4H-SiC.
It is initiated by an excess concentration of intrinsic defects, most likely
silicon interstitials as the analysis of recent experiments
indicates\cite{konstantinov:92,laube:99,bracht:00,janson:00} in contrast to
the earlier assumption\cite{mokhov:84} of a vacancy-mediated mechanism.
Similar mechanisms are operative in the more fundamental self diffusion.  As a
matter of fact, a microscopic understanding of the mechanisms underlying the
self diffusion should provide insight into the mechanisms of the dopant
diffusion.  In the most recent studies\cite{hong:79,hon:80,hong:80,hong:81} of self
diffusion in 3C- and 4H-SiC diffusion constants have been measured for the
carbon and silicon self diffusion. The basic diffusion mechanisms, however,
have not been unraveled.  In particular, the role of interstitials and
vacancies in different diffusion processes remains open for both polytypes.

Alongside with the diffusion, the annealing of mobile intrinsic defects
contains information about the migration of vacancies and interstitials.  The
annealing of vacancy-related defects has been studied by positron annihilation
spectroscopy (PAS)\cite{kawasuso:96,brauer:96,polity:99,kawasuso:01} and electron
paramagnetic resonance
techniques (EPR)\cite{itoh:97,wimbauer:97,vonbardeleben:00,son:01} in irradiated
material. The identification of the EPR-centers as isolated
silicon\cite{itoh:97,wimbauer:97} and carbon\cite{son:01} vacancies has been
verified theoretically.\cite{wimbauer:97,bockstedte:02} It was found that the
annealing behavior of carbon and silicon vacancies shows striking differences.
For example, in EPR-Experiments\cite{itoh:97} the silicon vacancy was found to
anneal in several stages at 150$^{\circ}$C, 350$^{\circ}$C and
750$^{\circ}$C, whereas for the carbon vacancy a single characteristic
annealing temperature of 500$^{\circ}$C was observed.\cite{son:01} Recently,
an EPR-center was observed by EPR experiments\cite{lingner:01} in
material annealed above 750$^{\circ}$C and interpreted as a vacancy-antisite
complex.  This complex may be a decay product of the metastable silicon
vacancy (in \emph{p}-type and intrinsic material) as predicted by
theory.\cite{rauls:00,bockstedte:00} A comprehensive interpretation of
annealing experiments in terms of elementary processes crucially depends on an
understanding of the underlying diffusion mechanisms of intrinsic defects.

\begin{figure}
\includegraphics[width=0.85\linewidth]{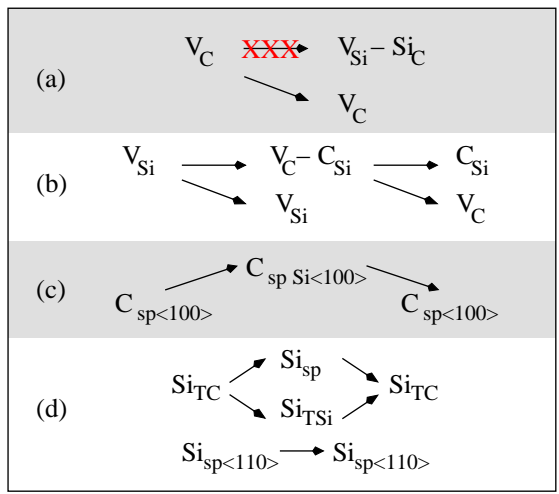}
\caption{\label{fig:scn.diff}The dominant migration mechanisms of 
  interstitials and vacancies: (a) carbon vacancies (V$_{\mathrm{C}}$)
  migrate on the carbon sublattice only, (b) transformation of the silicon
  vacancy (V$_{\mathrm{Si}}$) into a carbon vacancy-antisite complex
  V$_{\mathrm{C}}$-C$_{\mathrm{Si}}$ (\emph{p}-type and compensated) and the vacancy migration on the
  silicon sublattice (\emph{n}-type), (c) carbon interstitial migration via
  split-interstitial configurations C$_{\mathrm{sp}\langle100\rangle}$ and
  C$_{\mathrm{sp\/Si}\langle100\rangle}$, and (d) kick-out mechanism for the carbon
  coordinated silicon interstitial (Si$_{\mathrm{T\/C}}$) via
  split-interstitials (Si$_{\mathrm{sp}\langle100\rangle}$ and
  Si$_{\mathrm{sp}\langle110\rangle}$)(\emph{p}-type) and direct migration of the
  silicon split-interstitial Si$_{\mathrm{sp}\langle110\rangle}$ (compensated
  and \emph{n}-type material).}
\end{figure}

Theoretical investigations by \emph{ab initio} methods can provide important
insight in a microscopic picture of the self- and dopant-diffusion as well as the
annealing kinetics.  In the present paper we investigate the mobile
intrinsic defects and their migration mechanisms in 3C-SiC using a method
based on density functional theory. The Frenkel pair recombination and other
diffusion-controlled annealing mechanisms are treated
elsewhere.\cite{bockstedte:03a} Our investigation is based on recent
theoretical studies of the ground state properties of
vacancies\cite{zywietz:99,torpo:01} and antisites\cite{torpo:01} and addresses
interstitials and vacancy-antisite-complexes that are relevant for the
migration.  We show that different types of silicon interstitials, which
have entirely different migration paths, are important in \emph{p}-type and \emph{n}-type
material. The metastability of the silicon vacancy in \emph{p}-type material is
demonstrated to have strong implications for vacancy-assisted diffusion
mechanisms. The considered diffusion processes are summarized in
Fig.~\ref{fig:scn.diff}, which shows the most important migration channels for
vacancies and interstitials in 3C-SiC. We find that interstitials are
by far more mobile than vacancies and therefore play a prominent role in
diffusion. We also demonstrate that the diffusion mechanisms strongly depend on
the charge state of the mobile defects.

The outline of the paper is as follows.  Beginning with a description of the
method in Section~\ref{sec:dft} we turn to the ground state properties of
interstitials, vacancies and vacancy-antisite-complexes in Section \ref{sec:i} and
\ref{sec:v}.  The migration mechanisms of vacancies and interstitials are
analyzed in Section~\ref{sec:v.i.mig}. The role of interstitials and
vacancies in the self- and dopant-diffusion is discussed in
Section~\ref{sec:discuss}. A summary concludes the paper.

\section{Method}
\label{sec:dft}
We carried out first principles calculations using the plane wave
pseudopotential package \textsf{FHI96MD}~\cite{bockstedte:97a} within the
framework of density functional theory (DFT).\cite{hohenberg:64,kohn:65} The
local density approximation~\cite{perdew:81} (LDA) is employed for the
exchange-correlation functional and spin effects are included within the LSDA
where necessary.

\begin{figure}
\includegraphics[width=0.8\linewidth]{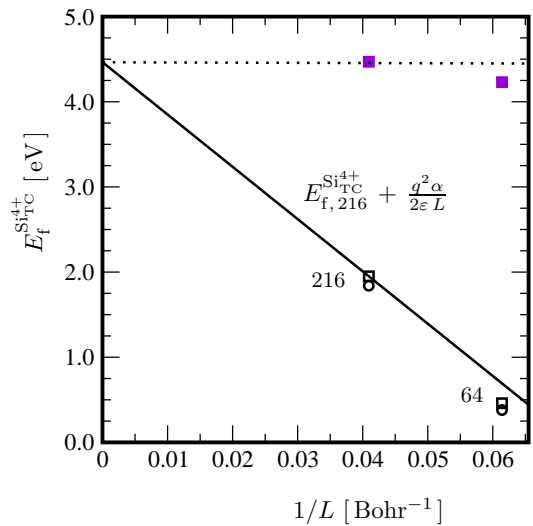}
\caption{\label{fig:madelung} Madelung corrections for the silicon interstitial
  Si$_{\mathrm{T\/C}}^{4+}$. The formation energy $E_{\mathrm{Si, f}}$ as
  calculated in the 64 and 216 atom cell with and without the Madelung
  correction (filled and open symbols, respectively) is plotted versus the
  inverse defect-defect distance. The squares and circles refer to a
  (2$\times$2$\times$2-mesh ) k-point set and the $\Gamma$-point. The
  extrapolation of $E_{\mathrm{f}}^{\mathrm{Si}_{\mathrm{T\/C}}}$ obtained in
    the 216 atom cell to the isolated defect using the monopole correction is shown by the solid line.}
\end{figure}

We describe the intrinsic defects and their environment using periodic
supercells.  Large supercells with 64 and 216 crystal lattice sites are used.
The electron transitions between the defect and its periodic images give rise
to the formation of defect bands and affect the defect energetics. This
artificial defect-defect interaction can be efficiently reduced by a special
k-point sampling.\cite{makov:96} A special k-point mesh~\cite{monkhorst:76}
with 8 k-points in the Brillouin zone (2$\times$2$\times$2-mesh) provides
converged defect energies for the 64-atom cell. It is crucial to preserve the
correct degeneracy of the isolated defect levels, as the degeneracy may lead
to a Jahn-Teller instability.  In those cases, where symmetry-lowering
Jahn-Teller-distortions are important, we perform the calculations using only
the $\Gamma$-point in a 216 atom cell (for which already sufficiently
converged formation energies are obtained), with the exact point symmetry of
an isolated defect.  In the case of charged defects we follow the approximate
procedure of Makov and Payne~\cite{makov:95} to account for the electrostatic
interaction of the periodically arranged defects as well as their interaction
with the compensating background. The relevance of such corrections has been
already pointed out by Torpo \emph{et~al.} for the vacancies in
4H-SiC.\cite{torpo:01} Especially for the highly charged interstitial defects
this procedure improves the defect energetics considerably.  In
Fig.~\ref{fig:madelung} we demonstrate this for the carbon-coordinated silicon
interstitial Si$_{\mathrm{T\/C}}$ (cf.  Sec.~\ref{sec:i}). We find that the
completely ionized interstitial silicon (charge state 4$^{+}$) is strongly
screened by the surrounding lattice. A comparison of the formation energies
calculated for the 64 and 216 atom cell with and without the Madelung
correction shows indeed that the correction leads to a more consistent
description. Whereas we obtain 0.5\,eV and 1.9\,eV without the correction
using special k-points, the corrected results of 4.23\,eV and 4.47\,eV agree
to within 0.24\,eV. Here we have only included the dominant monopole correction
($q^{2}\/\alpha/2\varepsilon\/L$, where $\alpha$ is the Madelung constant of
the simple cubic lattice, $L$ the defect-defect distance and $\varepsilon$ the
experimental dielectric constant), the quadrupole correction ($\sim
q\/Q/\varepsilon\/L^{3}$, where $Q$ is the quadrupole moment) should lead to a
smaller correction as suggested by the detailed analysis of Lento
\emph{et~al.}\cite{lento:02}

Soft norm-conserving pseudopotentials of the Troullier-Martins
type~\cite{troullier:91} and a plane-wave basis set are employed. The carbon
pseudopotential has been optimized for calculations with a small basis set
(matching radii $r^{{\rm c}}_{{\rm s}}$=1.6\,Bohr, $r^{{\rm c}}_{{\rm
    p}}$=1.7\,Bohr and $r^{{\rm c}}_{{\rm d}}$=1.5\,Bohr). For the silicon
potential the standard parameters are used. With these pseudopotentials and a
basis set including plane waves of a kinetic energy up to 30\,Ry defect
formation energies have been calculated with a basis set induced convergence
error below 0.1\,eV (for an energy cut-off of 40\,Ry all energy differences
are converged).  For each defect the coordinates of all atoms in the supercell
have been relaxed. This is essential to obtain a correct energetic
description. 
\subsection{Formation energies}
As a guide through the hierarchy of intrinsic defects one often uses their
relative abundance in thermodynamic equilibrium. At constant volume the
concentration of defects is determined by the free energy of defect formation $F_{\mathrm{f}}$:
\begin{equation}
c_{{\rm D}}=c_{{\rm S}}\,\exp{\left(-\frac{F_{{\rm f}}}{k_{{\rm B}}\,T}\right)}\quad ,
\label{eq:cntrn}
\end{equation}
where $c_{{\rm S}}$ is the concentration of the sites that are open to the defect, $k_{{\rm
    B}}$ is the Boltzmann constant and $T$ is the temperature. The free
energy of formation is specified by the formation energy $E_{{\rm f}}$ and
the formation entropy $S_{{\rm f}}$ of the defect:
\begin{equation}
F_{{\rm f}}=E_{{\rm f}}-T\,S_{{\rm f}}\quad.
\end{equation}
Usually the defect abundance is well described by the formation energy
$E_{\mathrm{f}}$ alone.  Even though the formation entropy may amount up to
10\,$k_{{\rm B}}$ (values of about 8\,$k_{{\rm B}}$ have been recently
reported for vacancies in Si~\cite{bloechl:93} and GaAs~\cite{bockstedte:97}),
its contribution can be relevant only at fairly high temperatures and for
defects with comparable formation energies.

In the compound material SiC the equilibrium defect concentration depends on
the chemical environment which is characterized by the carbon and silicon
chemical potentials, $\mu_{{\rm C}}$ and $\mu_{{\rm Si}}$. The chemical
potentials may vary only within certain bounds to prevent the formation of
other more stable phases than SiC. In excess of carbon (C-rich conditions) such a phase can be
graphite. In the other extreme
(Si-rich conditions), this would be crystalline silicon. The formation energy of a defect is
calculated as~\cite{zhang:91,scherz:93} 
\begin{equation}
E_{{\rm f}}=E_{\rm D,\ cell}-n_{{\rm C}}\,\mu_{{\rm C}}-n_{{\rm
    Si}}\,\mu_{{\rm Si}}-n_{{\rm e}}\,\mu_{{\rm F}}\quad,
\label{eq:form.raw}
\end{equation}
where $E_{{\rm D,\ cell}}$ is the total energy of the defect supercell, $n_{{\rm
    C}}$ and $n_{{\rm Si}}$ are the number of carbon and silicon atoms in the
supercell. In case of an ionized defect, $n_{{\rm e}}$ is the number of
excess electrons transferred from the Fermi-level $\mu_{{\rm F}}$ to the
localized defect levels (for positively charged defects $n_{{\rm e}}$ is
negative). In equilibrium, the chemical potentials of the environment $\mu_{{\rm
    C}}$ and $\mu_{{\rm Si}}$ are related to the chemical potential of the crystal
$\mu_{\mathrm{SiC}}$ by $\mu_{\rm SiC}=\mu_{\rm C}+\mu_{\rm Si}$. With this
relation we eliminate $\mu_{{\rm C}}$ in Eq.~(\ref{eq:form.raw}). We express the remaining free parameter $\mu_{{\rm Si}}=\mu_{{\rm
    Si}}^{0}+\Delta\mu$ in terms of the chemical potential of crystalline
silicon $\mu_{{\rm Si}}^{0}$ and the difference to this value $\Delta\mu$.
These substitutions in Eq.~(\ref{eq:form.raw}) yield
\begin{equation}
E_{{\rm f}}^{n_{\rm e}}=E_{{\rm D}}^{n_{\rm e}}-(n_{{\rm Si}}-n_{{\rm
    C}})\,\Delta\mu-n_{{\rm e}}\,\mu_{{\rm F}}\quad,
\label{eq:form}
\end{equation}
where $E_{{\rm D}}^{n_{\rm e}}=E_{{\rm D,\ cell}}^{n_{\rm e}}-n_{{\rm
    C}}\,\mu_{{\rm SiC}}-(n_{{\rm Si}}-n_{{\rm C}})\,\mu_{{\rm Si}}^{0}$ is
expressed by the number of the involved atoms and chemical potentials, which
are obtained by total energy calculations for the perfect crystals. As
described above, neither $\mu_{{\rm C}}$ nor $\mu_{{\rm Si}}$ can exceed the
values of the corresponding bulk phases (graphite or diamond and silicon).
Therefore $\Delta\mu$ may vary only between the negative heat of formation of
the corresponding polytype $-H_{{\rm f,\ SiC}}$ and 0 (C-rich and Si-rich
conditions, respectively). Our calculated value of the heat of formation
amounts to 0.61\,eV and depends only slightly on the polytype. This result is
in good agreement with the value of 0.58\,eV obtained by Zywietz {\em et
  al.\/}~\cite{zywietz:99} and slightly lower than the experimental value of
0.72\,eV.  Here we have deliberately chosen diamond as the reference. This
choice is comparable to the alternative choice of graphite due to the small
energy difference between graphite and diamond. Note that our definition of
the formation energy $E_{{\rm D}}$ is independent of this aspect.  The value
of $E_{{\rm D}}$ of Ref.~\onlinecite{zywietz:99} is retained by evaluating
E$_{\mathrm{f}}$ for $\Delta\mu=-1/2\,H_{{\rm f,\ SiC}}$.  For the calculation
of charged defects it is convenient to state the value of the Fermi-level
$\mu_{{\rm F}}$ relative to the valence band edge $E_{{\rm V}}$. In this case
the energy $E_{{\rm D}}$ is given by the formula
\begin{equation}
E_{{\rm D}}^{n_{\rm e}}=E_{{\rm D,\ cell}}-n_{{\rm C}}\,\mu_{{\rm SiC}}-(n_{{\rm
    Si}}-n_{{\rm C}})\,\mu_{{\rm Si}}^{0}-n_{{\rm e}}\,E_{\rm V}\quad,
\end{equation}
The value of $E_{{\rm V}}$ is not reliably accessible
in the calculation of the involved defects. Therefore we take the value $E_{{\rm V}}$ from a
calculation for the perfect crystal and remove shifts in the average potential
by aligning the bulk-spectrum features in the density of states of the defect
cell and the defect-free cell. The alignment is consistent within 0.06eV
over the whole energy range. By this procedure we also verified that
the defect cells contain a sufficiently large bulk-like
environment around the defect. 

The realization of a specific charge state of a deep defect for a given
Fermi-level $\mu_{\mathrm{F}}$ requires that the formation energy of the
defect in the corresponding charge state is lower than for the other possible
charge states and that all occupied single particle levels lie below the
conduction band edge. Even though the Kohn-Sham levels have no physical
interpretation by the design of the theory (only the electron density and the
total energy are a well defined quantities), they reproduce the experimental
quasi-particle band structure quite well.  Yet, the calculated Kohn-Sham band
gaps are usually smaller than experimental values (in 3C-SiC and 4H-SiC we
obtain 1.2\,eV and 2.2\,eV as compared to the experimental findings of
2.39\,eV and 3.27\,eV). In this paper we follow a common practice (cf. e.g.
Ref.~\onlinecite{torpo:01}) and use the experimental value for the conduction
band edge. With this approach it is not possible to unambiguously determine
the charge states that are only stable in a region close to the conduction
band edge.

The thermodynamic ionization level of a charged defect is given by the value
of the Fermi-level at
which the defect alters its charge state. It is obtained from the formation
energy by
\begin{equation}
\varepsilon(q_{2}\vert q_{1})=E_{\rm D}^{q_{1}}-E_{\rm D}^{q_{2}}\quad.
\end{equation}
Here $q_{1}$ and $q_{2}$ indicate the different charge states of the defect.
Usually only one electron is transfered between the electron reservoir and the
defect levels. A simultaneous transfer of two electrons is unfavorable due to
the cost of the effective electron-electron repulsion. However, in some
cases~\cite{anderson:75,baraff:79} the electron-electron repulsion can be
compensated by a configurational relaxation arising from a strong
electron-phonon coupling. This effect, known as the negative-$U$ effect, leads to
an attractive effective electron-electron interaction and the ionization level
$\varepsilon(q-1\vert q)$ appears below $\varepsilon(q|q+1)$.

\subsection{Defect migration}
In order to analyze possible migration paths of interstitials and vacancies we
apply two different standard methods. For the interstitial migration we use an
implementation~\cite{pehlke:00p} of the ridge method of Ionova and
Carter.\cite{ionova:93} In this method, a saddle point search is conducted for
given initial and final configurations of the migration event, both
configurations being slightly distorted towards each other.  The two
configurations are iterated such that they approach each other and the energy barrier
along the line connecting the two configurations is minimized. By
this procedure both configurations converge to the saddle point and the lowest
migration energy barrier $E_{\mathrm{m}}$ is obtained. The search is repeated for all
relevant charge states and the obtained transition states are analyzed.  This
automatic search of the saddle point fails when defect levels of different symmetry cross
along the migration path. In this situation, which applies to the migration of
vacancies, we analyze the potential energy surface using the drag method. This
approach has been previously employed in a study of the gallium vacancy migration in
GaAs.\cite{bockstedte:97} For a relevant set of coordinates (reaction
coordinates), the potential energy surface is calculated by constraining
the coordinates of interest and minimizing the total energy for all remaining
coordinates.  The consistency of our choice of the reaction coordinates is
verified. This includes the smoothness of the potential energy surface as a
function of the reaction coordinates. Details of the procedure are given in
Sec.~\ref{sec:v.i.mig}, where the individual migration mechanisms are discussed.

\section{Interstitials}
\label{sec:i}
\subsection{Silicon interstitials}
\label{sec:i.si}
\begin{figure}
\vspace*{0pt}
\includegraphics[width=0.9\linewidth]{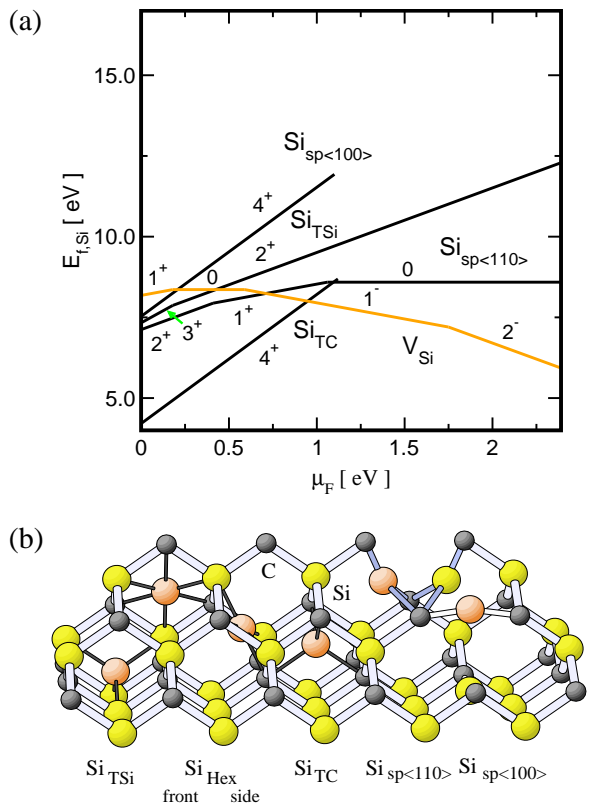}
\caption{Silicon-interstitials: (a) formation energy for Si-rich
  conditions in comparison to the silicon vacancy (b) geometry of
  investigated interstitial sites; the hexagonal
  site is not stable.}
\label{fig:isi}
\end{figure}

SiC as a compound semiconductor possesses various interstitial configurations,
some of them being distinguished only by a carbon- or silicon-like environment.
Important interstitial configurations are depicted in Fig.~\ref{fig:isi}.
There are tetrahedral configurations with either four silicon or carbon
neighbors and a hexagonal configuration. Besides these, split-interstitial
configurations and bond-center configurations can exist. In a split
configuration an interstitial silicon atom shares a site with a lattice atom,
which could be either carbon or silicon. These dumbbell-like  interstitials
occur with orientations of the atom pair in the $\langle100\rangle$- and the
$\langle110\rangle$-directions.  In a bond-center configuration the
interstitial silicon is centered at the bond of two neighboring lattice atoms.

We have considered all of the above configurations in our investigation. The
hexagonal and bond-center interstitials as well as the split-interstitials on
the carbon sublattice turned out to be unstable in all relevant charge states.
The calculated formation energies for the stable interstitials in comparison
with the formation energy of the silicon vacancy (c.f. Sec.~\ref{sec:v}) are
displayed in Fig.~\ref{fig:isi} as functions of the Fermi-level $\mu_{{\rm
    F}}$. To make the relation to Eq.~(\ref{eq:form}) more transparent, we
consider the case $\Delta\,\mu=0$, i.e.  Si-rich conditions, in
Figs.~\ref{fig:isi} and \ref{fig:ic}. Since in case of silicon interstitials
$n_{\mathrm{Si}}-n_{\mathrm{C}}=1$, their formation energies rise by 0.61\,eV
when referring to C-rich conditions, while the formation energy of the silicon
vacancy drops by the same amount. The charge states of the defects in the
different ranges of $\mu_{\mathrm{F}}$ are indicated. Except for the
split-interstitial Si$_{{\rm sp}\langle110\rangle}$ all silicon interstitials
are positively charged.

The hierarchy of silicon interstitials strongly depends on the position of the
Fermi-level. This is an immediate consequence of the existence of the deep
interstitial levels.  In \emph{p}-type material the most abundant is the
tetrahedrally carbon-coordinated interstitial Si$_{\rm T\/C}$ followed by the
split-interstitial Si$_{{\rm sp}\langle110\rangle}$. Around mid-gap, this
ordering changes: as the interstitial Si$_{{\rm T\/C}}$ is completely ionized
(carrying a charge of 4$^{+}$) its formation energy rises more rapidly than
that of the Si$_{{\rm sp}\langle110\rangle}$ interstitial, which is neutral
for $\mu_{\mathrm{F}}$ around mid-gap. For Si$_{\mathrm{T\/C}}$ we can only
estimate the formation energy of the charge states 3$^{+}$ (c.f.  below).
This also applies for the split-interstitial
Si$_{\mathrm{sp}\langle100\rangle}$. Therefore we have plotted the formation
energy only up to the estimated position of the lowest ionization level. The
tetrahedrally silicon-coordinated interstitial Si$_{\rm T\/Si}$ and the
split-interstitial with $\langle100\rangle$-orientation have a higher
formation energy than the other stable configurations.

\begin{figure}
\includegraphics[width=\linewidth]{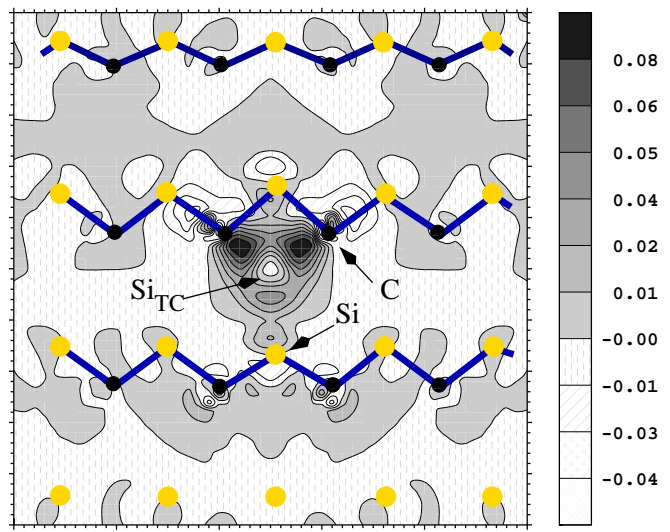}
\caption{Binding of Si$_{{\rm TC}}$: polarization
  $\delta\/n(\mathbf{r})$ of the electron density around the bare interstitial silicon ion at the
  Si$_{\mathrm{T\/C}}$-site obtained as a difference between the electron
  densities of the interstitial $n_{\mathrm{Si}_{\mathrm{T\/C}^{4+}}}$ and of
  the bulk crystal $n_{\mathrm{bulk}}$ (calculated in the
  216-atom cell).}
\label{fig:n.SiTC}
\end{figure}

In the following we describe the electronic structure of silicon
interstitials.  The interstitial Si$_{{\rm T\/C}}$ does not introduce deep
levels into the band gap. The silicon atom loses its valence electrons and
becomes completely ionized. However, the interaction with the carbon and
silicon neighbors induces an electron flow to the silicon interstitial,
resulting in an electron distribution that resembles that at a regular silicon
lattice site. The \emph{s}- and \emph{p}-components of the electron density integrated over
a sphere with a radius of 1.75\,Bohr (half the nearest neighbor distance in
SiC) around the Si-atom deviate from the values for a silicon site in a
perfect crystal by less than 1\%.  This is also true for spheres of smaller
radii.  In Fig.~\ref{fig:n.SiTC} we have plotted the charge redistribution
$\delta\,n=n_{\mathrm{Si}_{\mathrm{TC}}^{4+}}-n_{\mathrm{bulk}}$, where
$n_{\mathrm{Si}_{\mathrm{TC}}^{4+}}$ is the electron density of the supercell
containing the interstitial and $n_{\mathrm{bulk}}$ refers to the density of
the perfect crystal. The polarization of the surrounding carbon-silicon bonds
and the formation of bonds with the silicon interstitial is clearly visible.
Although we have not found any localized one-particle levels within the band
gap, we do not expect the charge state 4$^{+}$ to persist throughout the whole
experimental band gap. Indeed, slightly above the experimental conduction band
energy we observe a localized state. Due to the large charge state correction
the corresponding ionization level $\varepsilon(3^{+}|4^{+})$ lies
substantially lower, in a region somewhat above the experimental mid-gap. Thus
the charge state 4$^{+}$ should be stable at least up to the mid-gap position
of the Fermi-level. Yet, the quantitative determination of the ionization
level is not possible in DFT-LSDA calculations due to the well-known band gap
problem. The localized level is in resonance with the Kohn-Sham (and
experimental) conduction band. To obtain their relative position (which is
crucial for a correct occupation of levels) one would have to apply a
XC-discontinuity correction to the conduction band and the localized state,
which is well beyond the scope of the present work. This problem is less acute
in 4H-SiC, where the band gap is substantially wider. We
found~\cite{mattausch:04} that the carbon-coordinated Si-interstitial in
4H-SiC possesses a similar ionization level $\varepsilon(3^{+}|4^{+})$ and a
corresponding one-particle state that is located within the Kohn-Sham band
gap.

The screening of the silicon interstitial weakens the neighboring
carbon-silicon bonds.  This is manifested by an inward
relaxation of the carbon neighbor-shell towards the interstitial by 3.5\% of
an ideal bond length.  The energy associated with this relaxation amounts to
4.82\,eV. The relaxation of the first and second neighbor shell yields
4.37\,eV, whereas the inclusion of only the first neighbor shell results in a
gain of 0.34\,eV. This behavior shows the strong coupling between the induced
polarization and the relaxation.
 
\begin{table}
\caption{
\label{tab:chrg.ci}
Ionization levels of carbon and silicon interstitials with respect to the
  valence band edge $E_{{\rm V}}$ in eV. The calculations have
  been performed in a 64-atom cell using special k-points (C$_{{\rm
  sp}\langle100\rangle}$ and C$_{{\rm sp\/Si}\langle100\rangle}$: 216-atom cell and
  $\Gamma$-point to include spin-polarization and Jahn-Teller
  distortions). Note that values for C$_{{\rm
  sp}\langle100\rangle}^{0}$ and C$_{{\rm
  sp}\langle100\rangle}^{-}$ correspond to the tilted configuration.}
\begin{ruledtabular}
\begin{tabular}{cccccccc}
&Si$_{\mathrm{T\/Si}}$&Si$_{\mathrm{sp}\langle110\rangle}$&C$_{{\rm sp}\langle100\rangle}$&C$_{{\rm sp\/Si}\langle100\rangle}$&C$_{{\rm Hex}}$&C$_{{\rm
       T\/Si}}$&C$_{\mathrm{T\/C}}$\\
\hline\\
(2$^{+}|$3$^{+}$)&0.2&--  &--  &--  &--  &--  &--  \\
(1$^{+}|$2$^{+}$)&-- &0.4&0.6&0.4&0.9&0.6&1.4\\
(0$|$1$^{+}$)    &-- &1.1&0.8&0.7&1.3&1.1&1.8\\
(1$^{-}|$0)      &-- &-- &1.8&1.9&2.3&1.9&2.3\\
(2$^{-}|$1$^{-}$)&-- &-- &-- &2.3&-- &-- &-- \\
\end{tabular}
\end{ruledtabular}
\end{table}

The tetrahedral interstitial Si$_{{\rm TSi}}$, in contrast to Si$_{{\rm TC}}$,
has a deep non-degenerate level in the band gap below mid-gap. Its occupation
enables charge states between 4$^{+}$ and 2$^{+}$. However, due to the charge
state corrections, Si$_{{\rm TSi}}^{4+}$ is not realized. Only a single
ionization level (2$^{+}|$3$^{+}$) at 0.2\,eV is present (c.f.
Tab.~\ref{tab:chrg.ci}). The state has s-like character at the interstitial
silicon and p-like character at the fourth neighbor carbon-atoms, which have a
common bond with four silicon neighbors of Si$_{\mathrm{T\/Si}}$. There is
only a negligible amplitude at the first neighbor shell. The screened charge
of the silicon ion induces a coupled rehybridization and relaxation: while the
silicon neighbor shell relaxes outward (4.1\,Bohr Si-Si$_{{\rm TSi}}$
distance), the carbon second neighbor shell moves inward with a C-Si$_{{\rm
    TSi}}$ distance of 3.94\,Bohr.  Apparently, the unfavorable Si-Si bonds
makes this interstitial site less favorable than the Si$_{{\rm TC}}$-site.  This
explains also the observed instability of the hexagonal site, which would lie
on the line connecting adjacent Si$_{{\rm TSi}}$- and Si$_{{\rm TC}}$-sites:
The three-fold carbon coordination cannot counterbalance the energy costs of
the unfavorable silicon-silicon bond distance until the tetrahedrally
carbon-coordinated site is reached.

In the split-interstitials, as displayed in Fig.~\ref{fig:isi}, the two
silicon atoms that comprise the interstitial have a mutual distance of
4.1\,Bohr (Si$_{\mathrm{sp}\langle110\rangle}$) and 6.1\,Bohr
(Si$_{\mathrm{sp}\langle100\rangle}$).  The bonds of the
Si$_{\mathrm{sp}\langle110\rangle}$-interstitial are rearranged such that each
of the silicon atoms has a three-fold carbon coordination and a weak
$\sigma$-bond to its silicon partner. Within the band gap we find two
non-degenerate deep levels below mid-gap (approximately at $E_{{\rm
    V}}+0.4\,eV$ and $E_{{\rm V}}+0.6\,eV$) that possess anti-bonding character
between the two silicon atoms. There is also a level close to the conduction
band. Hence, one may expect that Si$_{\mathrm{sp}\langle110\rangle}$ could
exist in the charge states 4$^{+}$ to 2$^{-}$. However, according to our
calculations only the charge states 2$^{+}$, 1$^{+}$ and neutral are realized.
In the Si$_{\mathrm{sp}\langle100\rangle}$-interstitial both silicon atoms are
located in the interstitial 3.1\,Bohr away from the silicon lattice site.
This structure corresponds to a complex of two
Si$_{\mathrm{T\/C}}$-interstitials and a silicon vacancy. We do not find deep
states in the band gap and the screening is almost prefect.  According to our
calculations this interstitial is almost unstable. However, it is an important
intermediate configuration in the migration of the Si$_{{\rm
    TC}}$-interstitial.

\subsection{Carbon interstitials}
\label{sec:i.c}
\begin{figure}
\vspace*{0pt}
\includegraphics[width=\linewidth]{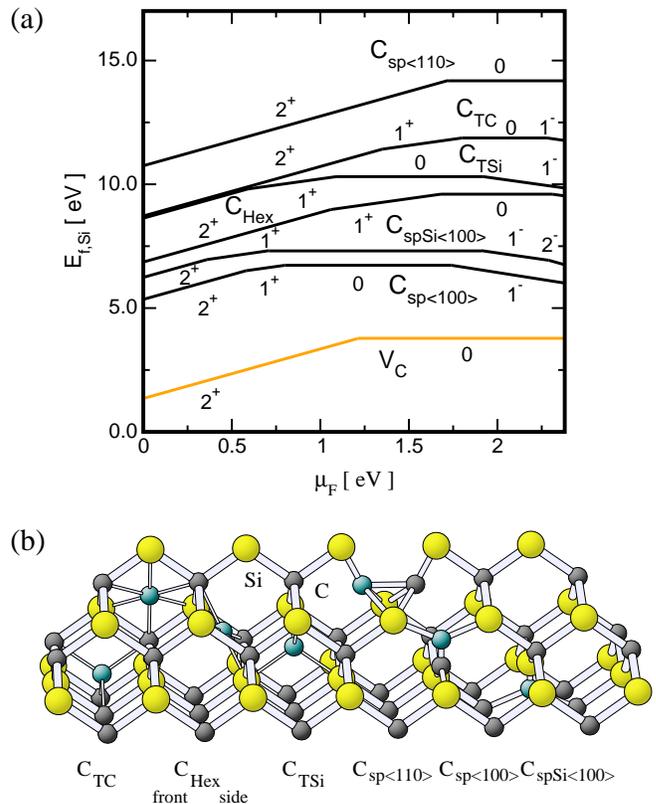}
\caption{\label{fig:ic}
Carbon-interstitials: (a) formation energy for Si-rich
  conditions in comparison to the carbon vacancy (b) geometry of
  investigated interstitial sites.}
\end{figure}

A different hierarchy is found for the carbon interstitials as for the silicon
interstitials. In Fig.~\ref{fig:ic} we have depicted the stable carbon
interstitial configurations.  The formation energies vs. the Fermi-level are
displayed in Fig.~\ref{fig:ic}a (for silicon rich conditions) in comparison
with the formation energy of the carbon vacancy (c.f. Sec.~\ref{sec:v}).  In
relation to Eq.~(\ref{eq:form}) we again refer to Si-rich conditions. For
C-rich conditions the formation energy of all carbon interstitials drops by
0.61\,eV, whereas that of the vacancy rises correspondingly. The
split-interstitials with $\langle100\rangle$-orientation (C$_{{\rm
    sp\langle100\rangle}}$ on the carbon sublattice and C$_{{\rm
    sp\/Si\langle100\rangle}}$ on the silicon sublattice ) have the lowest
formation energy under all doping conditions. The hexagonal interstitial
follows next in the hierarchy. The tetrahedral interstitials and the
split-interstitial with $\langle110\rangle$-orientation are less favorable.
For all interstitials deep levels exist in the band gap.
The outlined hierarchy reflects the preference of carbon to form short bonds in
the carbon split-interstitial configurations C$_{{\rm sp\langle100\rangle}}$
and C$_{{\rm sp\/Si\langle100\rangle}}$.

\begin{figure}
\includegraphics[width=0.8\linewidth]{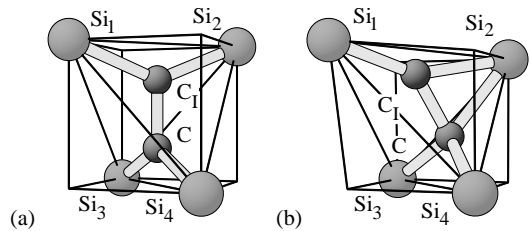}
\caption{\label{fig:geom.icsh}
Geometry of the carbon split-interstitial
C$_{\mathrm{sp}\langle100\rangle}$: (a) $\langle100\rangle$-orientation with
D$_{\mathrm{2d}}$-/D$_{2}$-symmetry, (b)
tilted C$_{\mathrm{sp}\langle100\rangle}$. The
C$_{\mathrm{sp\/Si}\langle100\rangle}$-configuration corresponds to the
untilted C$_{\mathrm{sp}\langle100\rangle}$ with the C-C$_{\mathrm{I}}$-pair
replaced by a Si-C$_{\mathrm{I}}$-pair and C-neighbors instead of Si-neighbors.
}
\end{figure}

For C$_{{\rm sp\langle100\rangle}}$ we find two different configurations
depending on the charge state, as illustrated in Fig.~\ref{fig:geom.icsh}. In the
positive charge states the axis of the carbon pair is oriented along the
$\langle100\rangle$ direction. In the neutral and negative charge states this
orientation becomes metastable. In the stable configuration the axis of the
carbon pair is tilted in the corresponding $\{011\}$-plane. The new stable
configuration arises from a pseudo Jahn-Teller distortion of the untilted
defect.

For C$_{\mathrm{sp}\langle100\rangle}^{2+}$ the carbon pair has a
$\langle100\rangle$-orientation.  The symmetry of the defect is
D$_{\mathrm{2\/d}}$.  We find a bond distance of 2.38\,Bohr within the
carbon-pair and a silicon-carbon bond distance of 3.53\,Bohr, which is close
to the undistorted bond distance in SiC. Along the $\langle100\rangle$-axis
the original tetrahedral arrangement of the silicon neighbors is substantially
distorted.  The bond angle between the carbon-silicon bonds amounts to
$137.5^\circ$.  The electronic structure is dominated by the sp$^{2}$-like
hybridization of the two carbon atoms. There is a two-fold degenerate level
($e$-representation) within the band gap.  Besides four carbon-silicon
$\sigma$-bonds, the localized wave function contains an unhybridized \emph{p}-orbital
at each carbon-atom which is oriented perpendicular to the plane spanned by
the $\sigma$-bonds.  In the case of C$_{\mathrm{sp}\langle100\rangle}^{+}$ the
partial occupation of the degenerate state invokes a Jahn-Teller distortion. The
distortion twists the defect molecule about its axis by 3.7$^{\circ}$ so that
the silicon neighbors leave their original positions in the \{011\}-planes.
This deformation lowers the energy as it allows a stronger interaction of the
carbon \emph{p}-orbital with the silicon orbital.  The symmetry becomes D$_{2}$.
Besides the positively charged defect also the neutral and negative defect can
be realized in this twisted configuration.  However, in the neutral charge
state the spin triplet is preferred over the singlet by 0.1\,eV and the
D$_{\mathrm{2\/d}}$-symmetry is retained. 

A tilting of the $\langle100\rangle$-axis by 31$^{\circ}$ (in
Fig.~\ref{fig:geom.icsh} in the plane Si$_{1}$-C$_{\mathrm{I}}$-Si$_{2}$)
transforms the metastable configuration of
C$_{\mathrm{sp}\langle100\rangle}^{0}$ and
C$_{\mathrm{sp}\langle100\rangle}^{-}$ into a stable configuration.  Alongside
with the tilt a simultaneous shift of the atom pair occurs, such that the distance
of the two carbon atoms to Si$_{\mathrm{2}}$ become similar (C$_{\mathrm{I}}$-Si$_{2}$:
3.7\,Bohr and C-Si$_{2}$: 3.6\,Bohr for C$_{\mathrm{sp}\langle100\rangle}^{0}$
and C$_{\mathrm{sp}\langle100\rangle}^{1-}$). While the distance within the
pair (2.48\,Bohr) only slightly increases, the distance
Si$_{1}$-C$_{\mathrm{I}}$ is reduced to 3.2\,Bohr. The distances Si$_{3}$-C
and Si$_{4}$-C remain unaffected.  The tilt and shift introduce a
rehybridization of the bonds. Besides the carbon-carbon bond, the carbon atom
C now binds to three silicon neighbors Si$_{2}$, Si$_{3}$ and Si$_{4}$, while
C$_{\mathrm{I}}$ maintains bonds with only two silicon neighbors -- a stronger
one with Si$_{1}$ and a weaker one with Si$_{2}$. The rebonding yields an energy gain of 0.6\,eV. The metastable $\langle100\rangle$-oriented
configuration is separated from the tilted ground state configuration by a
low energy barrier. In the neutral state the estimated  energy
barrier is only 0.06\,eV.

The split-interstitial C$_{{\rm sp\/Si\langle100\rangle}}$ has a C$_{\rm
  2v}$-symmetry. Hence, for this interstitial only non-degenerate levels occur
within the band gap, which have a similar sp$^{2}$-character as C$_{{\rm
    sp\langle100\rangle}}$. There are three levels around mid-gap. The lowest
one has $\sigma$-character. The two higher ones have a bonding character
with the carbon neighbors and an admixed \emph{p}-character on the carbon atom.
C$_{{\rm sp\/Si\langle100\rangle}}$ occurs in the charge states 2$^{+}$ to
2$^{-}$, with the lowest electronic level being always completely occupied
(c.f. the ionization levels in Tab.~\ref{tab:chrg.ci}).  The bond length of
the atom pair is slightly shorter than the ideal SiC-bond.  This is also true
for the bond between the silicon atom of the pair and its other carbon
neighbors. The carbon-carbon bonds with a length of 2.8\,Bohr are considerably
shorter. Note that Jahn-Teller distortions are absent in this case. However,
with the occupation of the non-degenerate defect levels the position of
the silicon-carbon pair shifts along its axis.

The hexagonal interstitial C$_{{\rm hex}}$ as reflected by its C$_{{\rm
    3v}}$-symmetry has a non-degenerate level above mid-gap and a two-fold
degenerate level below the conduction band. The states originate from a
localized \emph{p}-orbital that is oriented along the symmetry axis and from weak
sp$^{2}$-like bonds with the carbon atoms of the hexagonal ring.  In the
relevant charge states 2$^{+}$, 1$^{+}$ and 0 only the non-degenerate
\emph{p}-like level is occupied. As a result the interstitial carbon is
shifted along the hexagonal axis out of the center of the ring towards the
carbon neighbors (bond distance 3.04\,Bohr as compared to the distance to the
silicon neighbors of 3.7\,Bohr). In the charge state 1$^{-}$ the occupation of
the degenerate level leads to Jahn-Teller distortion, that lifts the axial
symmetry. As a result, the interstitial carbon approaches two of the carbon
neighbors.
    
A similar behavior is observed for the tetrahedrally carbon-coordinated
configuration (C$_{\mathrm{T\/C}}$), which has a non-degenerate and a
three-fold degenerate level within the band gap.  Yet, the three-fold
coordinated configuration is energetically preferred over the tetrahedral
configuration, since in the latter case the optimization of the carbon-carbon
bonds results in an unfavorable elongation of surrounding carbon-silicon
bonds.
At the silicon-coordinated site favorable carbon-carbon bonds are not available.
Therefore these configurations as well as the $\langle 110\rangle$-oriented
split configuration are scarcely occupied in the thermodynamic equilibrium.

\section{Vacancies and vacancy-antisite-complexes}
\label{sec:v}
\subsection{Vacancies}
The properties of the carbon and silicon vacancies (V$_{\mathrm{C}}$ and
V$_{\mathrm{Si}}$) in 3C- and 4H-SiC have been discussed in detail in the
literature.~\cite{wimbauer:97,torpo:98,deak:99,zywietz:99,zywietz:00,zywietz:00a,torpo:01}
We summarize these results as necessary ingredients for the following discussion of the
vacancy-antisite complexes and the mechanisms of the vacancy
migration.

In Figures \ref{fig:isi} and \ref{fig:ic} our results for the formation energy
of V$_{\mathrm{Si}}$ and V$_{\mathrm{C}}$ are compared to the formation energy
of the corresponding interstitials for Si-rich conditions. V$_{\mathrm{C}}$ is
by far more abundant than carbon and silicon interstitials. This dominance
also prevails in stoichiometric and C-rich material.  V$_{\mathrm{Si}}$, on
the other hand, is less abundant than the interstitials in \emph{p}-type material and
dominates under intrinsic and \emph{n}-type conditions.  Furthermore, it has a higher
formation energy than carbon vacancies. Hence, the equilibrium concentration
of silicon vacancies is by several orders of magnitude lower.

The electronic structure of both unrelaxed vacancies derives from the four
dangling bonds of the nearest neighbors. They have T$_{{\rm d}}$-symmetry and
possess a three-fold degenerate one-electron level ($t_{2}$-representation)
within the band gap. The corresponding non-degenerate $a_{1}$-level falls
below the valence band edge.  The occupation of the degenerate $t_{2}$-level
either induces a Jahn-Teller distortion or leads to the formation of a
non-degenerate high-spin state.  The particular distinction between
V$_{\mathrm{C}}$ and V$_{\mathrm{Si}}$ is the extended character of the
silicon dangling bonds of the former versus the strong localization of carbon
dangling bonds of the latter.~\cite{zywietz:99} Driven by the overlap between
the silicon dangling bonds, a considerable Jahn-Teller distortion of
V$_{\mathrm{C}}$ is observed.~\cite{zywietz:99,torpo:01} For
V$_{\mathrm{Si}}$, in contrast, correlation effects are more
important~\cite{wimbauer:97,torpo:98,deak:99} than the electron-phonon
coupling. As a consequence, a multiplet state is
formed~\cite{deak:99,zywietz:00a} which leads to the preference for a
high-spin states,\cite{wimbauer:97,torpo:99} as predicted by
DFT-calculations~\cite{torpo:99,zywietz:99} for all charge states.

In Tab.~\ref{tab:chrg.vccsi} we have summarized the ionization levels of the
two vacancies in 3C- and 4H-SiC for the cubic and hexagonal site. In the
following we judge the realization of the different charge state using the
experimental band gap as noted in Sec.~\ref{sec:dft}. According to our
findings, V$_{\mathrm{Si}}$ exists in the charge states 1$^{+}$ through
2$^{-}$ in 3C-SiC. In 4H-SiC also the charge states 3$^{-}$ and 4$^{-}$ are
possible.  V$_{\mathrm{Si}}^{2+}$ is not realized as a ground state with the
inclusion of charge state corrections, a finding~\cite{mattausch:01} that is
verified by Ref.~\onlinecite{torpo:01}.  Our ionization levels are in good
agreement with the results of Ref.~\onlinecite{torpo:01} obtained including
charge state corrections. A slight deviation from the results of Zywietz
\emph{et~al.}\cite{zywietz:99} originates from their neglect of these
corrections and a different definition of the reference for the negative
charge states. The realization of the charge states 3$^{-}$ and 4$^{-}$ is
also observed in DFT-LSDA calculation in 6H-SiC\cite{lingner:01} using the
LMTO-ASA Greens function approach as well as the charge state $2^{+}$. In
comparison with our results in 4H-SiC, the calculated ionization levels agree
to within 0.3\,eV including the levels $\varepsilon(3^{-}|2^{-})$ and
$\varepsilon(4^{-}|3^{-})$.  We consider this as a good agreement taking the
variation of the levels among the different polytypes and the inequivalent
lattice sites into account (it was not stated in Ref.~\onlinecite{lingner:01}
to which site the calculations referred).

The carbon vacancy exists only in the charge states 2$^{+}$ and neutral in
3C-SiC. An enhanced electron-phonon coupling found for V$_{\mathrm{C}}^{0}$
and V$_{\mathrm{C}}^{2-}$ destabilizes the charge states 1$^{+}$ and 1$^{-}$.
The negative-\emph{U} amounts to 0.15\,eV (V$_{\mathrm{C}}^{+}$). However, as the
recent identification of V$_{\mathrm{C}}^{+}$ by a comparison of the
experimental hyperfine tensors and the calculated hyperfine interaction in
4H-SiC\cite{bockstedte:02} indicates, the prediction of the small negative-\emph{U}
maybe artifical.  In 4H-SiC also the charge states 1$^{-}$ and 2$^{-}$ can be
realized apart from the negative-\emph{U} effect found for V$_{\mathrm{C}}^{-}$.
Note that the ionization levels of V$_{\mathrm{C}}$ vary between the cubic and
hexagonal site.  This stems from the different arrangement of the third
nearest neighbors and translates into a different hybridization of the defect
levels.
\subsection{Vacancy-antisite-complexes and the metastability of V$_{\mathrm{Si}}$}
\begin{figure*}
\includegraphics[width=\linewidth]{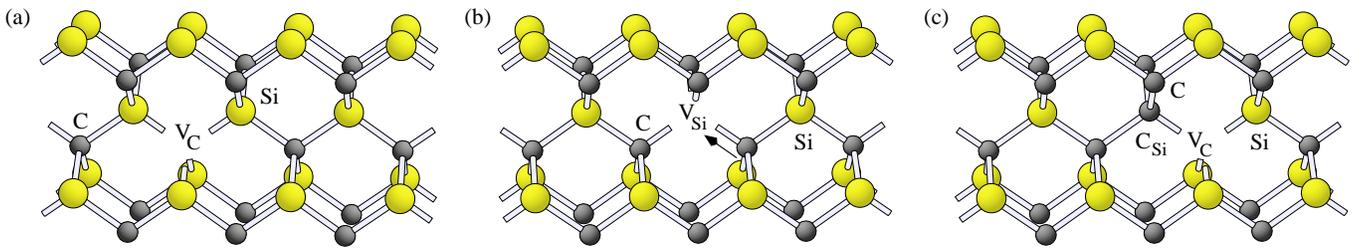}
\caption{\label{fig:geom.v} Geometry of (a) the carbon vacancy V$_{\mathrm{C}}$, (b) the
  silicon vacancy V$_{\mathrm{Si}}$ and (c) the carbon
  vacancy-antisite-complex V$_{\mathrm{C}}$-C$_{\mathrm{Si}}$. The hop of a
  carbon neighbor that transforms the silicon vacancy into a carbon
  vacancy-antisite-complex is indicated by an arrow in (b).}
\end{figure*}
\begin{figure}
\includegraphics[width=\linewidth]{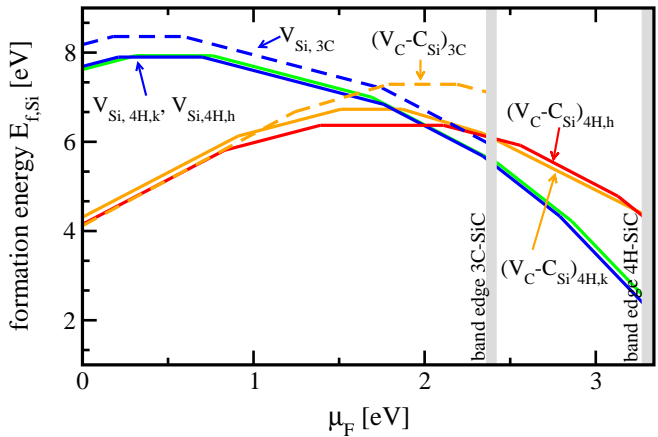}
\caption{\label{fig:energy.vsi.vccsi}Formation energy of the silicon vacancy V$_{\mathrm{Si}}$
  and the carbon vacancy-antisite-complex V$_{\mathrm{C}}$-C$_{\mathrm{Si}}$
  for 3C-SiC and 4H-SiC. The experimental band edges of 3C- and 4H-SiC are indicated by
  vertical bars.}
\end{figure}
For the migration two vacancy-antisite complexes are important:
V$_{\mathrm{Si}}$-Si$_{\mathrm{C}}$ and V$_{\mathrm{C}}$-C$_{\mathrm{Si}}$,
where the antisite is the nearest neighbor of the vacancy. These complexes 
are obtained from V$_{\mathrm{C}}$ and V$_{\mathrm{Si}}$, by a displacement of
a silicon/carbon neighbor into the empty lattice site as illustrated in
Fig.~\ref{fig:geom.v}.

We have found that the V$_{{\rm Si}}$-Si$_{{\rm C}}$ complex does not possess a
stable configuration in all relevant charge state. As a result it decays into the
carbon vacancy. The origin is a size effect: both, the silicon antisite and the
silicon vacancy, show a strong outward relaxation.\cite{torpo:99} For the
antisite this relaxation optimizes the silicon-silicon bond distance to its
neighbors, yet the bond remains strained as the ideal value of silicon bulk
cannot be achieved. In the vacancy-antisite complex the missing silicon neighbor
gives room for further relaxation by moving
the antisite towards the vacancy. Additionally, the energy of the dangling bond
located at the antisite is reduced by interacting with the carbon neighbors of
the vacancy. As a result the barrier towards the vacant site vanishes. The
unstable vacancy-antisite complex transforms into a stable carbon vacancy.
From the above arguments we expect that our result for
3C-SiC~\cite{bockstedte:00} is also valid in other polytypes. This expectation
is supported by a DFT based tight-binding calculation for the neutral complex
in 4H-SiC.\cite{rauls:00}

In contrast to V$_{\mathrm{Si}}$-Si$_{\mathrm{C}}$ the carbon vacancy-antisite
complex V$_{{\rm C}}$-C$_{{\rm Si}}$ is stable. In fact, it is even more
stable than the silicon vacancy in \emph{p}-type material and for a Fermi-level in a
mid-gap range. This is shown in Fig.~\ref{fig:energy.vsi.vccsi}. Before we discuss the details and
consequences of this metastability of the silicon vacancy, we briefly
turn to the electronic structure of the complex.

Similar to the vacancies its localized levels within the band gap derive from
the four dangling bonds at the carbon antisite and the silicon neighbors of
the vacancy. By the C$_{3\/\mathrm{v}}$-symmetry of the ideal defect these
dangling bonds give rise to two non-degenerate $a$-levels and a doubly
degenerate $e$-level. As for the vacancies one $a$-level forms a resonance in the
valence band. The other $a$-level lies at E$_{\mathrm{V}}$+1.8\,eV and the
$e$-level is found at E$_{\mathrm{V}}$+2.1\,eV
((V$_{\mathrm{C}}$-C$_{\mathrm{Si}}$)$^{{+}}$).  Whereas the character of the
localized $a$-levels is dominated by the carbon dangling bond at the antisite,
the $e$-level derives from the three silicon dangling bonds. In the
charge states 2$^{+}$, 1$^{+}$ and 0 the $a$-level is occupied. The
occupation of the $e$-level for (V$_{\mathrm{C}}$-C$_{\mathrm{Si}}$)$^{{-}}$
gives rise to a Jahn-Teller effect, which lowers the symmetry to
C$_{\mathrm{1\/h}}$. According to our findings, the charge state
(V$_{\mathrm{C}}$-C$_{\mathrm{Si}}$)$^{2-}$ is 
unstable (c.f.  Tab.~\ref{tab:chrg.vccsi}).  In 4H-SiC the carbon
vacancy-antisite complex shows similar properties. For our present purpose it
is sufficient to treat the complexes aligned along the hexagonal axis. In this
case the vacancy and the antisite occupy either neighboring hexagonal or cubic sites.
The larger band gap of 4H-SiC --3.3\,eV as compared to 2.4\,eV in 3C-SiC--
enables the negative charge states 2$^{-}$ and 3$^{-}$.
\begin{table}
\caption{\label{tab:chrg.vccsi} Ionization levels of the native vacancies and
  the carbon vacancy-antisite complex in 3C- and 4H-SiC (cubic and hexagonal
  sites are denoted as k and h respectively). The results have been obtained using a
  216 atom cell (3C) and a 128 atom cell including spin-polarization for the
  silicon vacancy.}
\begin{ruledtabular}
\begin{tabular}{llcccccc}
&&$(+\vert2^{+})$&$(0\vert+)$&$(-\vert0)$&$(2^{-}\vert-)$&$(3^{-}\vert2^{-})$&$(4^{-}\vert3^{-})$\\
\hline
V$_{\mathrm{C}}$&3C&1.29&1.14&2.69&2.04&--&--\\
V$_{\mathrm{C}}$&4H,\,k&1.29&1.17&2.23&2.11&--&--\\
V$_{\mathrm{C}}$&4H,\,h&1.05&0.97&2.23&2.05&--&--\\
\hline
V$_{\mathrm{Si}}$&3C   &--&0.18&0.61&1.76&--&--\\
                 &4H,\,k&--&0.21&0.70&1.76&2.34&2.79\\
                 &4H,\,h&--&0.30&0.76&1.71&2.42&2.90\\
\hline
V$_{\mathrm{C}}$-C$_{\mathrm{Si}}$&3C   &1.24&1.79&2.19&--&--&--\\ 
V$_{\mathrm{C}}$-C$_{\mathrm{Si}}$&4H,\,k&0.91&1.51&1.87&2.34&3.3&--\\
V$_{\mathrm{C}}$-C$_{\mathrm{Si}}$&4H,\,h&0.83&1.39&2.11&2.56&3.12&--\\

\end{tabular}
\end{ruledtabular}
\end{table}

The metastability of V$_{\mathrm{Si}}$ and its stabilization in \emph{n}-type
material occurs in both 3C- and 4H-SiC. As
Fig.~\ref{fig:energy.vsi.vccsi} shows, V$_{{\rm C}}$-C$_{{\rm Si}}$ is more
stable than V$_{\mathrm{Si}}$ under \emph{p}-type conditions and for a mid-gap
Fermi-level. Its transformation into V$_{{\rm C}}$-C$_{{\rm Si}}$ is
associated with a large energy gain (about 4\,eV for a Fermi-level at the
valence band). Since the defect levels of the
V$_{\mathrm{C}}$-C$_{\mathrm{Si}}$-complex are located above the levels
of V$_{\mathrm{Si}}$, the energy gain is reduced by the successive occupation
of defect levels. This leads to a stabilization of  V$_{\mathrm{Si}}$ over 
V$_{{\rm C}}$-C$_{{\rm Si}}$ for \emph{n}-type conditions. Only a metastability was
observed~\cite{mattausch:01} using special k-points in a 64 atom cell. In this case,
the formation energy of the silicon vacancy is higher in the negative charge
states than the value we obtain in the 216 atom cell.  The silicon vacancy and the
vacancy-antisite complex have similar formation energies for a Fermi-level
above mid-gap in 4H-SiC, for the cubic complex at $\mu_{\mathrm{F}}$=1.8\,eV
and for the hexagonal complex at $\mu_{\mathrm{F}}$=2.0\,eV. In 3C-SiC the
formation energies match at $\mu_{\mathrm{F}}$=1.8\,eV.  The strong
interaction between the vacancy and the antisite is reflected in the binding
energy of the complexes. The binding energy ranges between 1.2\,eV (2$^{+}$)
and 1.0\,eV (1$^{-}$).

Using a LMTO-ASA Greens function approach based on DFT and LSDA
Lingner~\emph{et al.}~\cite{lingner:01} obtain similar results for 6H-SiC.  A
metastability of the silicon vacancy is also predicted by their calculations.
The Fermi-level at which the V$_{\mathrm{Si}}$ and the complex have similar
energies is located above 2.8\,eV. However, in their calculations the
metastability is more pronounced which leads to a larger energy gain of the
transformation.  According to Fig. 7 of Ref.~\onlinecite{lingner:01} the
energy difference amounts to 4\,eV (2.6\,eV for the unrelaxed defects) for the
neutral defects. Using DFT-based tight-binding calculations the authors obtain
1.6\,eV. Our calculations yield 1.1\,eV for 3C-SiC, 1.1 and 1.3\,eV for cubic
and hexagonal defects in 4H-SiC, which is much closer to the tight binding
result.  For the carbon vacancy-antisite complex they obtain similar
ionization levels for the positive and neutral charge states of the unrelaxed
complex. Including relaxations a negative-$U$ effect is found that is not
observed in our calculations.  Yet, as noted above the agreement of the
ionization levels of the silicon vacancy is good. It indicates that the
formation energy of the silicon vacancy is essentially shifted by about 3\,eV
(1.3\,eV for the unrelaxed defects) with respect to the value of the carbon
vacancy-antisite complex as compared to our calculations.  Consequently, the
Fermi-level at which the silicon vacancy and the complex have a comparable
formation energy is found closer to the conduction band edge as in our case.
\section{Migration}
\label{sec:v.i.mig}
\subsection{carbon vacancies}
\begin{figure}
\includegraphics[width=0.65\linewidth]{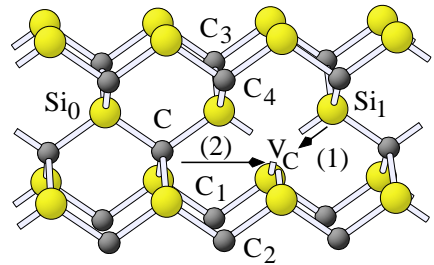}
\caption{\label{fig:mig.vc} Migration of the carbon vacancy V$_{\mathrm{C}}$:
  (1) initial nearest neighbor hop, (2) migration on the carbon sublattice by
      second neighbor hops.}
\end{figure}

For the migration of the vacancies we have analyzed two mechanisms that
involve either nearest neighbor hops as observed in silicon~\cite{bloechl:93}
or second neighbor hops, that are e.g. relevant for the gallium vacancy in
GaAs.\cite{chen:94,dabrowski:94,bockstedte:97} The two hops are depicted in
Fig.~\ref{fig:mig.vc}.  As discussed in Sec.~\ref{sec:v}, nearest neighbor
hops transform the vacancy into a vacancy-antisite complex. In successive hops
the vacancy would proceed leaving behind a chain of antisites, which is
energetically unfavorable. In the ring mechanism proposed by Van
Vechten~\cite{vanvechten:84} this is avoided by the vacancy passing twice
through the same six-fold ring.  We have shown in Sec.~\ref{sec:v} that the
silicon vacancy-antisite complex is unstable. For the carbon vacancy, this
finding rules out any migration mechanism based on nearest neighbor hops.

In our analysis of the migration by second neighbor hops and the
calculation of the migration barriers, we follow the procedure in
Ref.~\onlinecite{bockstedte:97}. There it was shown for the gallium vacancy in
GaAs that the migration mechanism involves the motion of the hopping gallium atom
(a C-atom in Fig.~\ref{fig:mig.vc}) and
its common neighbor with the vacancy (the Si-neighbor in
Fig.~\ref{fig:mig.vc}). At the saddle point, the hopping atom passes through the plane
spanned by the four atoms (C$_{1}$,\ldots,C$_{4}$ in Fig.~\ref{fig:mig.vc})
that we refer to as the gate below.
In order to accommodate the strain of this configuration it is favorable for
the nearest neighbor to relax towards either of the vacant lattice sites. Thus
two saddle points and correspondingly two migration paths exist for the hop --
one, where the neighbor of the vacancy passes through the gate prior to the
second neighbor and one where the latter passes first.  Following
Ref.~\onlinecite{bockstedte:97}, we introduce two reaction coordinates
$\xi_{\mathrm{C}}$ and $\xi_{\mathrm{Si}}$ for the hopping carbon atom and the
common silicon neighbor by
\begin{equation}
\xi_{\mathrm{C/Si}}=\left(\mathbf{R}_{\mathrm{C/Si}}-\frac{1}{4}\sum_{i=1}^{4}\,\mathbf{R}_{\mathrm{C}_{i}}\right)\cdot
\mathbf{e}_{110}\quad,
\end{equation}
where $\mathbf{R}_{\mathrm{C/Si}}$ for $\xi_{\mathrm{C}}$ and
$\xi_{\mathrm{Si}}$, respectively, refers to the
coordinates of the carbon or silicon atom and the
$\mathbf{R}_{\mathrm{C}_{i}}$ are the coordinates of the four carbon atoms
comprising the gate; $\mathbf{e}_{110}$ is the unit vector of the
direction connecting the initial and final lattice site of the vacancy (here
the (110)-direction is chosen). 
\begin{table}
\caption{\label{tab:v.mig}Vacancy migration: the migration barrier and geometry at
      the transition state for the migration by second neighbor hops 
      and the transformation of V$_{\mathrm{Si}}$ into
      V$_{\mathrm{C}}$-C$_{\mathrm{Si}}$ 
      (initial nearest neighbor hop). See the text for details.}
\begin{ruledtabular}
\begin{tabular}{clccccc}
&&2$^{+}$&1$^{+}$&0&1$^{-}$&2$^{-}$\\
\hline
V$_{\mathrm{C}}$$\rightarrow$V$_{\mathrm{C}}$&
E$_{\textrm{m}}$ [\,eV\,]&5.2&4.1&3.5&-&-\\
& $\xi_{\textrm{Si}}$ [\,Bohr\,]&0.82&0.79&0.78&-&-\\
&Si-Si$_{0}$ [\,Bohr\,]&4.80&4.79&4.76&-&-\\
&Si-Si$_{1}$ [\,Bohr\,]&7.07&6.79&6.57&-&-\\
\hline
V$_{\mathrm{Si}}$$\rightarrow$V$_{\mathrm{Si}}$&E$_{\textrm{m}}$ [\,eV\,]&-&3.6&3.4&3.2&2.4\\
&$\xi_{\textrm{C}}$ [\,Bohr\,]&-&0.57&0.30&0.01&0.03\\
&C-C$_{0}$ [\,Bohr\,]&-&5.08&5.56&5.93&5.97\\
&C-C$_{1}$ [\,Bohr\,]&-&6.74&6.33&5.96&6.00\\
\hline
V$_{\mathrm{Si}}$$\rightarrow$V$_{\textrm{C}}$-C$_{\textrm{Si}}$&
E$_{\mathrm{m}}$ [\,eV\,]&-&1.9&2.4 (2.2)&2.5&2.7\\
V$_{\textrm{C}}$-C$_{\textrm{Si}}$$\rightarrow$V$_{\mathrm{Si}}$&
E$_{\mathrm{m}}$ [\,eV\,]&6.1&4.2&3.5&2.4&-\\
 &$\Delta\xi_{\mathrm{NN}}$ [\,Bohr\,]&1.96&1.53&1.64 (1.97)&1.67&1.67\\
 &$\xi_{\mathrm{NN},\perp}$ [\,Bohr\,]&0.00&0.05&0.19 (-0.11)&0.12&0.28\\
\end{tabular}
\end{ruledtabular}
\end{table}
Instead of evaluating the full potential energy surface for these two
coordinates, we consider only the region in the vicinity of the transition
state. For a given $\xi_{\mathrm{C}}$ the total energy is minimized with
respect to $\xi_{\mathrm{Si}}$ and the other atomic coordinates. We find
indeed the behavior we have discussed above. The transition state lies by
0.2\,eV lower than the configuration where the carbon atom and its silicon
neighbor pass through the gate at the same time. The barriers are obtained
using the 216 atom cell together with the $\Gamma$-point, which also allowed
for the proper inclusion of Jahn-Teller distortions in the initial
configurations. The results for the migration barrier and the geometry at the
transition state are given in Tab.~\ref{tab:v.mig}. Using a 64 atom cell and
special k-points ($2\times2\times2$ Monkhorst-Pack mesh), the calculated
barriers agree within 0.3\,eV. We find relatively high migration barriers
between 5\,eV (V$_{\mathrm{C}}^{2+}$), 4.1\,eV (V$_{\mathrm{C}}^{+}$) and
3.5\,eV (V$_{\mathrm{C}}^{0}$).  The migration barrier in \emph{n}-type material and
with a Fermi-level around mid-gap (V$_{\mathrm{C}}^{0}$) is much smaller than
in \emph{p}-type material (V$_{\mathrm{C}}^{2+}$). This stems from the different
occupation of bonding defect levels at the transition state. Besides the bond
between the common Si-neighbor and Si$_{\mathrm{0}}$ (c.f. Fig.~\ref{fig:mig.vc}), the remaining
three dangling bonds at the opposite vacant site contribute to the defect levels in
the band gap. With the occupation of the lowest level the corresponding
neighbor (Si$_{1}$) relaxes towards the vacant site, thus reducing the distance
to the silicon neighbor and the migration barrier.
\subsection{silicon vacancies}
In case of the silicon vacancy the migration by second neighbor hops competes
with the transformation
V$_{\mathrm{Si}}\,\leftrightarrow\,$V$_{\mathrm{C}}$-C$_{\mathrm{Si}}$.
Particularly in \emph{p}-type material, the metastability of the silicon vacancy will
have consequences for its migration.  However, a migration of the silicon
vacancy solely based on nearest neighbor hops is impossible. This is due to
our finding that the second configuration in such a mechanism, the
V$_{\mathrm{Si}}$-Si$_{\mathrm{C}}$-C$_{\mathrm{Si}}$-complex, is unstable in
all relevant charge states.  In order to understand the consequences of the
metastability, the barriers for the second neighbor hop and the transformation
V$_{\mathrm{Si}}\,\leftrightarrow\,$V$_{\mathrm{C}}$-C$_{\mathrm{Si}}$ have to
be compared.

The analysis of the second neighbor hop proceeds as detailed in the
previous section. This mechanism has essentially the same features as in the
case of the carbon vacancy. Now the second silicon neighbor and the
nearest carbon neighbor take the role of the respective carbon and silicon
atom in the discussion of V$_{\mathrm{C}}$.  The calculations have been
performed within the 216-atom cell using the $\Gamma$-point and including spin
polarization. For the 64-atom cell and a special k-point sampling we obtain
within 0.2\,eV the same barriers. Our results for the barriers and the
geometry at the transition state are listed in Tab.~\ref{tab:v.mig}. At the
transition state the displacement of the carbon neighbor out off the gate is
less pronounced as in the case of the carbon vacancy. For the negative vacancy it
almost vanishes. We explain this by the rather small extend of the carbon
orbitals and the resulting weak interaction of the carbon neighbor with the
other carbon dangling bonds. The migration barrier for this mechanism varies between
3.6\,eV (V$_{\mathrm{Si}}^{+}$) and 2.4\,eV (V$_{\mathrm{Si}}^{2-}$).

Our analysis of the nearest neighbor hop parallels the approach to the second
neighbor hop. We introduce the reaction coordinate
$\xi_{\mathrm{NN}}$ which measures the distance of the hopping carbon atom
from the center of mass of its three silicon neighbors. It is given by 
\begin{equation*}
\xi_{\mathrm{NN}}=\left(\mathbf{R}_{\mathrm{C}}-\frac{1}{3}\sum_{i=1}^{3}\,\mathbf{R}_{\mathrm{Si}_{i}}\right)\cdot\,\mathbf{e}_{111}\quad,
\end{equation*}
where $\mathbf{R}_{\mathrm{C}}$ is the coordinate of the hopping carbon
atom and $\mathbf{R}_{\mathrm{Si}_{i}}$ are the coordinates
of its three silicon neighbors; $\mathbf{e}_{111}$ is the unit vector in the
direction of the hop (here chosen as the (111)-direction). 
Note that this approach allows for a migration
path that deviates from the (111)-direction.

As detailed in Sec.~\ref{sec:v}, V$_{\mathrm{Si}}$ and
\mbox{V$_{\mathrm{C}}$-C$_{\mathrm{Si}}$} are found in different charge states
for a given Fermi-level. We assume that the charge state of the defect is
preserved during the transformation.  The necessary electron
transfer to obtain the equilibrium charge state occurs in the final
configuration.  Harrison has shown~\cite{harrison:98} for the case of the
interstitial in silicon that this assumption is reasonable.  This applies also
to the transformation of
V$_{\mathrm{Si}}^{2-}\,\rightarrow\,$(V$_{\mathrm{C}}$-C$_{\mathrm{Si}}$)$^{2-}$ and
(V$_{\mathrm{C}}$-C$_{\mathrm{Si}}$)$^{2+}\,\rightarrow\,$V$_{\mathrm{Si}}^{2+}$,
where the final configurations (V$_{\mathrm{C}}$-C$_{\mathrm{Si}}$)$^{-}$ and
V$_{\mathrm{Si}}^{+}$ are the ground state configurations. 

The energy barriers and the values of the reaction coordinate
$\xi_{\mathrm{NN}}$ at the transition state are listed in
Tab.~\ref{tab:v.mig}.  The barrier for the transformation
V$_{\mathrm{Si}}\,\rightarrow\,$V$_{\mathrm{C}}$-C$_{\mathrm{Si}}$ moderately
depends on the charge state. It is by 0.8\,eV lower for the positive vacancy
than for the doubly negative charge state. The reverse transformation
V$_{\mathrm{C}}$-C$_{\mathrm{Si}}\,\rightarrow\,$V$_{\mathrm{Si}}$ shows a
much stronger charge state dependence. A barrier of 6.1\,eV is found for
(V$_{\mathrm{C}}$-C$_{\mathrm{Si}}$)$^{2+}$, that lowers down to 2.4\,eV for
(V$_{\mathrm{C}}$-C$_{\mathrm{Si}}$)$^{-}$.  During the transformation the
hopping carbon atom moves away from the (111)-direction.  In this way it
optimizes the number of bonds maintained with its silicon and carbon
neighbors. A path along the (111)-direction has a much larger energy barrier,
due to a simultaneous breaking of three bonds with the silicon neighbors. Only
this latter path was obtained\cite{mattausch:01} using the 64 atom cell with a
special k-point sampling and was reproduced using the 216 atom cell. However,
this path is not important.

For all charge states the transformation
V$_{\mathrm{Si}}\rightarrow$V$_{\mathrm{C}}$-C$_{\mathrm{Si}}$ mainly occurs
in a low-spin state of the vacancy. Typically much higher barriers are found
for the high-spin state, except for V$_{\mathrm{Si}}^{0}$ where the high-spin
state has a slightly lower barrier (given in braces in Tab.~\ref{tab:v.mig}).
Note that the ground state of the vacancy-antisite complex is always a
low-spin state for all charge states.  Thus the reverse transformation
V$_{\mathrm{C}}$-C$_{\mathrm{Si}}\rightarrow$V$_{\mathrm{Si}}$ should occur
into the low-spin state.

In equilibrium (with a given Fermi-level), V$_{\mathrm{Si}}$ and
V$_{\mathrm{C}}$-C$_{\mathrm{Si}}$ are in different charge states. Provided
the temperature is sufficiently low and the transformation is a rare
event, the charge state can equilibrate at the final position before the
reverse process takes place. Only at extreme temperatures we expect
this picture to break down. Consequently, energy barriers for
different charge states should apply for the two processes.  In \emph{p}-type and
compensated material the barrier for the transformation
V$_{\mathrm{Si}}\,\rightarrow\,$V$_{\mathrm{C}}$-C$_{\mathrm{Si}}$ is with
1.9\,eV (V$_{\mathrm{Si}}^{1+}$) to 2.4\,eV (V$_{\mathrm{Si}}^{-}$) much lower
than the barrier of the reverse transformation of 6.1\,eV
((V$_{\mathrm{C}}$-C$_{\mathrm{Si}}$)$^{2+}$, $\mu_{\mathrm{F}}<1.24$\,eV) to
3.5\,eV
((V$_{\mathrm{C}}$-C$_{\mathrm{Si}}$)$^{0}$,$\mu_{\mathrm{F}}<2.19$\,eV).  In
\emph{n}-type material (i.e. for $\mu_{\mathrm{F}}>2.19$) this situation is reversed,
when (V$_{\mathrm{C}}$-C$_{\mathrm{Si}}$)$^{-}$ becomes the ground state. Now the
transformation of (V$_{\mathrm{C}}$-C$_{\mathrm{Si}}$)$^{-}$ into
V$_{\mathrm{Si}}^{-}$ has a higher probability than the opposite process
V$_{\mathrm{Si}}^{2-}\rightarrow$(V$_{\mathrm{C}}$-C$_{\mathrm{Si}}$)$^{2-}$.

A comparison between the migration barriers of the silicon vacancy and the
transformation barriers should tell whether the silicon vacancy is trapped as
a V$_{\mathrm{C}}$-C$_{\mathrm{Si}}$-complex or whether it freely migrates.
In \emph{n}-type material V$_{\mathrm{Si}}^{2-}$ is stable (i.e. for
$\mu_{\mathrm{F}}>1.76$\,eV). The migration barrier (2.4\,eV) is slightly
smaller than the transformation barrier (2.7\,eV) to the metastable
V$_{\mathrm{C}}$-C$_{\mathrm{Si}}$-complex. For $\mu_{\mathrm{F}}>2.19$\,eV
(V$_{\mathrm{C}}$-C$_{\mathrm{Si}}$)$^{-}$ returns by a reverse transformation
(2.4\,eV) via V$_{\mathrm{Si}}^{-}$ to V$_{\mathrm{Si}}^{2-}$ with an even
higher probability.  For
$\mu_{\mathrm{F}}$ between 1.79\,eV and 2.19\,eV
(V$_{\mathrm{C}}$-C$_{\mathrm{Si}}$)$^{0}$ is relevant. Even though the
barrier of the transformation
V$_{\mathrm{C}}$-C$_{\mathrm{Si}}\rightarrow$V$_{\mathrm{Si}}$ (3.5\,eV) is
larger, the migration of V$_{\mathrm{Si}}$ is essentially not hindered as
V$_{\mathrm{Si}}^{2-}$ is the ground state.  Thus, in \emph{n}-type material, the
vacancy should freely migrate.

On the other hand, for a Fermi-level below 1.7\,eV the
V$_{\mathrm{C}}$-C$_{\mathrm{Si}}$-complex is the ground state. Before a
migration of V$_{\mathrm{Si}}$ may take place, it has to be preceeded by a
transformation V$_{\mathrm{C}}$-C$_{\mathrm{Si}}\rightarrow$V$_{\mathrm{Si}}$.
This gives rise to an additional barrier. For a Fermi-level below mid-gap this
barrier becomes large, i.e.  4.1\,eV for
(V$_{\mathrm{C}}$-C$_{\mathrm{Si}}$)$^{+}$ and 6.1\,eV for
(V$_{\mathrm{C}}$-C$_{\mathrm{Si}}$)$^{2+}$. This additional barrier allows a
migration of V$_{\mathrm{C}}$-C$_{\mathrm{Si}}$ via V$_{\mathrm{Si}}$ only at
elevated temperatures, when the transformation
V$_{\mathrm{C}}$-C$_{\mathrm{Si}}\rightarrow$V$_{\mathrm{Si}}$ is thermally
activated. At low temperatures this migration is kinetically hindered.

Instead, another migration channel is opened: the
V$_{\mathrm{C}}$-C$_{\mathrm{Si}}$ complex can dissociate at the expense of
the binding energy. However, using the migration barrier for V$_{\mathrm{C}}$
between 3.5\,eV (V$_{\mathrm{C}}^{0}$) to 5\,eV (V$_{\mathrm{C}}^{2+}$) plus
the binding energy as an estimate for the dissociation barrier, we
conclude that the dissociation and the reverse transformation have similar
probabilities.  In \emph{p}-type and compensated material ($\mu_{F}<1.71$\,eV) the
migration of silicon vacancies is suppressed for moderate temperatures.

\subsection{carbon-interstitials}
Apparently, the energetically lowest sites are the most relevant ones for the migration
of the carbon interstitials. As discussed in Sec.~\ref{sec:i}, these are the
two split-interstitials C$_{\textrm{sp} \langle 100 \rangle}$ and
C$_{\textrm{spSi} \langle 100 \rangle}$.  The migration of the energetically
more favorable C$_{\textrm{sp} \langle 100 \rangle}$-interstitial may proceed
by two alternative routes: (a) by nearest neighbor hops between adjacent carbon
and silicon lattice sites in the sequence C$_{\textrm{sp} \langle 100
  \rangle}\rightarrow$C$_{\textrm{spSi} \langle 100
  \rangle}\rightarrow$C$_{\textrm{sp} \langle 100 \rangle}$ and (b) by second
neighbor hops of the interstitial between neighboring carbon lattice sites.

\begin{figure*} \includegraphics[width=\linewidth]{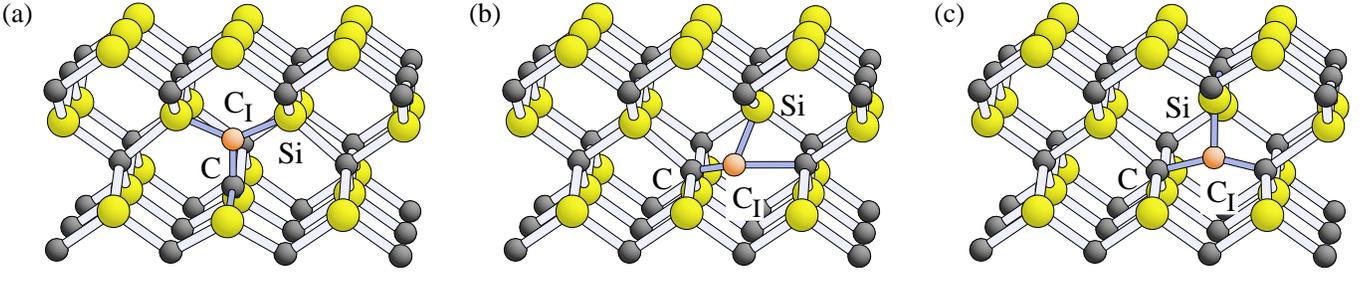}
\caption{\label{fig:mig.ci}Migration of carbon interstitials: (a) initial
  interstitial C$_{\mathrm{sp}\langle100\rangle}$, (b) transition state, (c)
  final interstitial C$_{\mathrm{sp\/Si}\langle100\rangle}$}
\end{figure*}

We illustrate the migration event by nearest neighbor hops in
Fig.~\ref{fig:mig.ci}. Taking C$_{\textrm{sp} \langle 100 \rangle}$
(Fig.~\ref{fig:mig.ci}a) as the initial configuration, the final configuration
is C$_{\textrm{spSi} \langle 100 \rangle}$ (Fig.~\ref{fig:mig.ci}c), which
is located at the nearest neighbor silicon site. The saddle point search
yields the transition state shown in Fig.~\ref{fig:mig.ci}b. The migration
proceeds by a concerted motion of the dumbbell, such that one carbon atom
approaches the neighboring silicon atom while the other returns into the
position on the lattice. At the same time, the silicon neighbor moves out of
this site to give room for a new carbon-silicon dumbbell.  We have conducted a
saddle point search in the charge states 2$^{+}$, 1$^{+}$, neutral and 1$^{-}$
using a 64 atom cell and a special k-point set. For the hop from
$\textrm{C}_{\textrm{sp}\langle 100 \rangle}$ to
$\textrm{C}_{\textrm{spSi}\langle 100 \rangle}$ we find a barrier of 1.7\,eV and
0.9\,eV for the charge states 2$^{+}$ and 1$^{+}$.  The barriers of the second
hop from $\textrm{C}_{\textrm{spSi}\langle 100 \rangle}$ to
$\textrm{C}_{\textrm{sp}\langle 100 \rangle}$ amount to 0.7\,eV and 0.2\,eV
respectively and are lower than those of the first hop. In the neutral and
negative charge state the hop starts from the tilted configuration. The path
is as discussed above.  The barriers amount to 0.5\,eV and 0.2\,eV for
C$_{\mathrm{sp}\langle100\rangle}$. Similar barriers are found in the case of
C$_{\mathrm{sp}\langle100\rangle}^{-}$.

The migration of C$_{\textrm{sp} \langle 100 \rangle}$ by second neighbor hops
can proceed in several different ways. Each of the carbon atoms of the
dumbbell is surrounded by six neighboring carbon lattice sites. While one atom
migrates to one of the neighboring carbon sites, the other moves back onto the lattice.
We have investigated the hop of the (001)-oriented split-interstitial in
the (110)-direction.  The migration proceeds in a similar manner as the nearest
neighbor hop, except that now the destination is the adjacent carbon lattice
site.  The migration barriers we obtained with the saddle point search are
listed in Tab.~\ref{tab:c.mig}. They are comparable to the nearest neighbor hop.
Only for C$_{\mathrm{sp}\langle100\rangle}^{2+}$ we find a lower barrier for
the second nearest neighbor hop (1.4\,eV) than for the nearest neighbor hop
(1.7\,eV)
\begin{table}
\caption{\label{tab:c.mig} Migration of carbon interstitials: Migration barriers
  $E_{\mathrm{m}}$ in eV for nearest neighbor hops
  C$_{\mathrm{sp}\langle100\rangle}\leftrightarrow$C$_{\mathrm{sp\/Si}\langle100\rangle}$
  and second neighbor hops. The migration
  of C$_{\mathrm{sp}\langle100\rangle}^{0}$ and C$_{\mathrm{sp}\langle100\rangle}^{-}$  starts from the tilted configuration.}
\begin{ruledtabular}
\begin{tabular}{lccccccc}
& 2$^{+}$&&1$^{+}$&&0&&1$^{-}$\\
\cline{2-2}\cline{4-4}\cline{6-6}\cline{8-8}
C$_{\mathrm{sp}\langle100\rangle}\rightarrow$
C$_{\mathrm{sp\/Si}\langle100\rangle}$&1.7&&0.9&&0.5&&0.7\\
C$_{\mathrm{sp\/Si}\langle100\rangle}\rightarrow$
C$_{\mathrm{sp}\langle100\rangle}$&0.9&&0.2&&0.2&&0.1\\[1ex]
C$_{\mathrm{sp}\langle100\rangle}\leftrightarrow$
C$_{\mathrm{sp}\langle100\rangle}$&1.4&&0.9&&0.5&&0.6\\
\end{tabular} 
\end{ruledtabular}
\end{table}

The relatively small migration barrier contrasts the much larger energy
difference found between the split-interstitials and the tetrahedrally
coordinated sites. As discussed in Sec.~\ref{sec:i.c}, at these
sites carbon bonds are established at high energy costs. 
Consequently, a relevant migration of carbon interstitials via these sites
is unlikely.

The hexagonal interstitial
is 1.4\,eV (C$_{\mathrm{Hex}}^{2+}$) and 1.6\,eV (C$_{\mathrm{Hex}}^{+}$) higher in energy than
$\textrm{C}_{\textrm{spSi}\langle 100 \rangle}$. This energy difference
is only slightly lower than the migration barrier for
$\textrm{C}_{\textrm{spSi}\langle 100 \rangle}^{2+}$ and even higher in the
case of  $\textrm{C}_{\textrm{spSi}\langle 100 \rangle}^{1+}$. Therefore we expect a
higher energy barrier for a migration via the hexagonal sites, particularly
as the mechanism should initially proceed similar to the migration via
split-interstitial sites to avoid extensive bond breaking. 

\subsection{silicon-interstitials}
In Section~\ref{sec:i.si} we have shown that the two most favorable interstitial
configurations are the tetrahedrally carbon-coordinated site and the
(110)-oriented split-interstitial. The first configuration is favored in
\emph{p}-type material and the latter is clearly dominating in \emph{n}-type material. 
The migration between the energetically lowest
sites may involve the other configuration as an intermediate site or may proceed by direct
hops. As discussed previously, we assume that the charge state is not altered
during a hop. Only in the intermediate configurations a charge transfer is allowed. 

\begin{figure*}
\includegraphics[width=\linewidth]{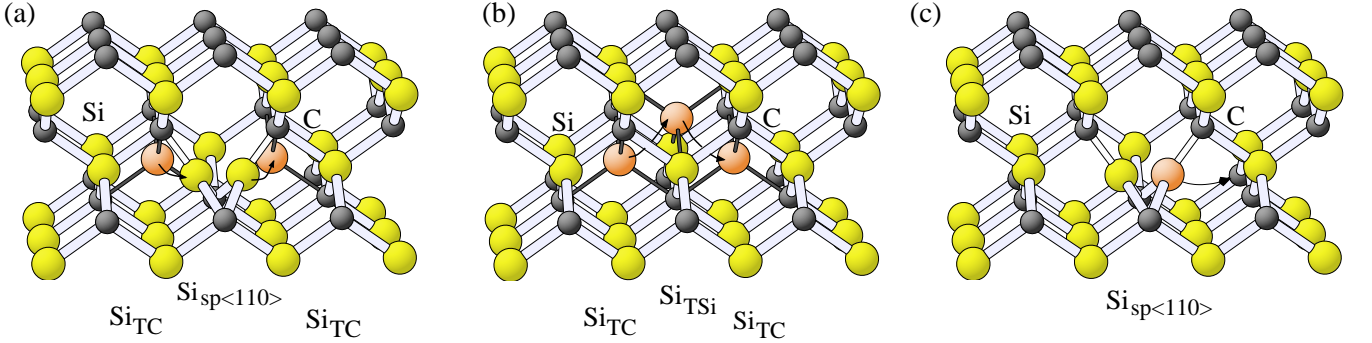}
\caption{\label{fig:mig.sii}Migration of silicon interstitials: (a) kick-out migration of
  Si$_{\mathrm{T\/C}}$ via the split-interstitial
  Si$_{\mathrm{sp}\langle110\rangle}$, (b) migration via the tetrahedral
  interstitial sites Si$_{\mathrm{T\/C}}$ and Si$_{\mathrm{T\/Si}}$, and (c) jumps
  between adjacent Si$_{\mathrm{sp}\langle110\rangle}$ configurations.}
\end{figure*}

In \emph{p}-type material all migration paths begin and end at the
Si$_{\mathrm{T\/C}}$ site. Two different types of migration mechanisms are
discussed here and depicted in Fig.~\ref{fig:mig.sii}: (a) a kick-out
mechanism that proceeds via the split-interstitials
Si$_{\mathrm{sp}\,\langle100\rangle}$ and Si$_{\mathrm{sp}\langle110\rangle}$
as intermediate sites and (b) a direct migration between the tetrahedral
interstitial sites Si$_{\mathrm{T\/C}}$ and Si$_{\mathrm{T\/Si}}$. The
kick-out mechanism consists of two steps. In the first step the tetrahedral
interstitial moves towards a silicon neighbor and kicks it out of its
lattice site. A split-interstitial is formed as an intermediate
site. Next, the second atom proceeds towards the closest tetrahedral interstitial site.
We have determined the migration paths and the energy barriers for the two
processes by a saddle point search using a 64-atom cell and special k-points
($2\times2\times2$ Monkhorst-Pack mesh). While the path via the
Si$_{\mathrm{sp}\,\langle100\rangle}$-configuration proceeds entirely along
the (100)-direction, the path via the Si$_{\mathrm{sp}\,\langle110\rangle}$
interstitial is curved. The transition state in both cases lies close to the
split-interstitial configuration. The energy barriers for the kick-out and the
kick-in process are listed in Tab.~\ref{tab:si.mig}.
\begin{table}
\caption{\label{tab:si.mig} Migration of silicon interstitials:
  Migration barriers $E_{\mathrm{m}}$ in eV for different migration paths. The
  migration paths are shown in Fig.~\ref{fig:mig.sii}. The charge state is
  4$^{+}$ unless indicated otherwise. See text
  for details.}
\begin{ruledtabular}
\begin{tabular}{cclclc}
&{T\/C}$\leftrightarrow$sp$\langle100\rangle$
&&T\/C$\leftrightarrow$sp$\langle110\rangle$
&&T\/C$\leftrightarrow$T\/Si \\
\cline{2-2}\cline{4-4}\cline{6-6}
1$\rightarrow$2&3.5&&3.4&&3.5\\
2$\rightarrow$1&0.03&&0.05&&0.3\\[2ex]
&sp$\langle110\rangle$ (+)&&sp$\langle110\rangle$
(0)&&sp$\langle110\rangle$ (0)\\
&$\leftrightarrow$\,sp$\langle110\rangle$&&$\leftrightarrow$\,sp$\langle110\rangle$&&$\leftrightarrow$\,sp$\langle1\bar{1}0\rangle$\\
\cline{2-2}\cline{4-4}\cline{6-6}
1$\leftrightarrow$2&1.0&&1.4&&1.3\\[2ex]
&T\/Si (2+)&&T\/Si (3+)&&T\/Si (2+)\\
&$\leftrightarrow$sp$\langle100\rangle$&&$\rightarrow$T\/C&&$\rightarrow$T\/C\\
\cline{2-2}\cline{4-4}\cline{6-6}
1$\rightarrow$2&0.1&&0.04&&0.03\\
2$\rightarrow$1&0.25&&-&&-\\
\end{tabular} 
\end{ruledtabular}
\end{table}
The kick-out barrier via the Si$_{\mathrm{sp}\,\langle100\rangle}$
interstitials amounts to 3.5\,eV and is only 0.1\,eV higher than the barrier
obtained via the Si$_{\mathrm{sp}\,\langle110\rangle}$ interstitial. The
results show that the two intermediate interstitial configurations are not
very stable as the barriers for the second event amount to 0.03\,eV and 0.05\,eV,
respectively. We expect that the interstitial passes through the intermediate
site without any significant delay. Therefore a charge transfer that
could convert the
Si$_{\mathrm{sp}\/\langle110\rangle}$-interstitial from the transient 4$^{+}$-state into 2$^{+}$
or 1$^{+}$ (the ground state in \emph{p}-type material) seems unlikely.

The migration between the tetrahedral interstitial sites Si$_{\mathrm{T\/C}}$
and Si$_{\mathrm{T\/Si}}$ proceeds directly through the interstice. The
migration path passes through the hexagonal configurations along the axis
connecting the initial and final configuration. The transition state lies
close to the ideal position of the hexagonal site. The saddle point search yields an
energy barrier of 3.5\,eV for the hop
Si$_{\mathrm{T\/C}}\rightarrow$\,Si$_{\mathrm{T\/Si}}$. For the hop Si$_{\mathrm{T\/Si}}\rightarrow$\,Si$_{\mathrm{T\/C}}$ we obtain a barrier
of 0.3\,eV. 

Since for Si$_{\mathrm{T\/Si}}$ the charge state 4$^{+}$ is not the ground
state, a charge transfer can change it with some probability. Then the
interstitial could alternatively migrate via the split-interstitial
Si$_{\mathrm{sp}\,\langle110\rangle}^{2+}$ with a low barrier (c.f.
Tab.~\ref{tab:si.mig}). However, this route is not very likely. Instead the
Si$_{\mathrm{T\/Si}}^{2+}$ (or Si$_{\mathrm{T\/Si}}^{3+}$) would move to the
Si$_{\mathrm{T\/C}}$-site with an even lower barrier.

Incompensated and \emph{n}-type  material, we consider the kick-out
migration via the split-interstitials Si$_{\mathrm{sp}\langle110\rangle}^{0}$.
As one of the silicon atoms of Si$_{\mathrm{sp}\langle110\rangle}$ hops
towards the next silicon site it kicks out the silicon atom at this site in
the corresponding (110)-direction to form a new split-interstitial. For this
process we find a migration barrier of 1.4\,eV.  For the positive
split-interstitial we obtain a barrier of 1.0\,eV. The orientation of the
split-interstitial can change by a rotation around the
$\langle100\rangle$-axis.  For example, the rotation around the (001)-axis
changes the orientation from (110) to (1$\bar{1}$0). It is associated with an
energy barrier of 1.3\,eV (Si$_{\mathrm{sp}\langle110\rangle}^{0}$).

In order to understand whether the kick-out migration via
Si$_{\mathrm{sp}\langle110\rangle}^{0}$ is the dominating mechanism, we have to
compare its activation energy with that of other possible mechanisms at
intrinsic and \emph{n}-type doping conditions.  In comparison to a migration via the
interstitial Si$_{\mathrm{T\/Si}}^{2+}$, we find that this configuration for
$\mu_{\mathrm{F}}$ above 1.3\,eV already lies higher in energy than the saddle
point of Si$_{\mathrm{sp}\langle110\rangle}$. Furthermore, according to our
estimates also Si$_{\mathrm{T\/C}}$ and Si$_{\mathrm{sp}\langle100\rangle}$
should be correspondingly higher in energy for a Fermi-level close to the
conduction band. Thus we conclude that the dominating mechanism under
intrinsic and \emph{n}-type conditions is the migration via the
Si$_{\mathrm{sp}\langle110\rangle}$-interstitial.

\section{Discussion}
\label{sec:discuss}
Our analysis of the mobile intrinsic point defects in the Sections \ref{sec:i}
and \ref{sec:v} and their migration mechanisms in Section \ref{sec:v.i.mig}
reveals a complex dependence of the abundance and the migration on the doping
conditions.  In the following we discuss the implications of our findings for
the self- and dopant-diffusion. The mechanisms that enter the discussion are
summarized schematically in Fig.~\ref{fig:scn.diff} and the associated energy
barriers are listed in Tabs.~\ref{tab:v.mig}, \ref{tab:c.mig}  and \ref{tab:si.mig}.  This
microscopic picture of the point defect migration has also implications for
the annealing mechanism of vacancies. This aspect is treated, together with an
analysis of Frenkel pairs, in a separate article.\cite{bockstedte:03a}
\subsection{Self diffusion}
The self diffusion constants of carbon
and silicon, $D_{\mathrm{C}}$ and $D_{\mathrm{Si}}$, are given by
the sum of different diffusion mechanisms, in particular of mechanisms
based on vacancy and interstitial migration. For instance, the carbon self diffusion
constant  is given by
\begin{equation}
D_{\mathrm{C}}=\sum_{i}D_{i}^{\mathrm{V}_{\mathrm{C}}}+\sum_{j}D_{j}^{\mathrm{I,\,C}}+\ldots
\end{equation}
where $D_{i}^{\mathrm{V}_{\mathrm{C}}}$ and $D_{j}^{\mathrm{I,\,C}}$ are the
diffusion constants of vacancy and interstitial mechanisms, respectively, and
$i$, $j$ denote the different migrations paths of the mobile defect in a
specific charge state. Each contribution is proportional to the concentration
of the mobile defect $c_{i}$ (c.f. Eq.~(\ref{eq:cntrn})) and its diffusivity
$d_{i}$, as given by the formation energy $E_{i}^{\mathrm{f}}$ and the
migration energy barrier $E_{i}^{\mathrm{m}}$, respectively,
\begin{equation}
D_{i}=D_{i}^{0}\,\exp{\left(-\frac{E_{i}^{\mathrm{f}}(\Delta\/\mu,\mu_{\mathrm{F}})+E_{i}^{\mathrm{m}}}{k_{\mathrm{B}}\/T}\right)}\quad,
\end{equation}
where $D_{i}^{0}$ is a constant prefactor that is determined by geometrical and
entropy factors. The activation energy
$Q_{i}=E_{i}^{\mathrm{f}}(\Delta\/\mu,\mu_{\mathrm{F}})+E_{i}^{\mathrm{m}}$ that
determines the exponential factor essentially characterizes the contribution
of each migration mechanism. The factor may vary over several orders of
magnitude for different mechanisms, whereas the prefactors $D_{i}^{0}$
typically have a similar order of magnitude.  Thus a comparison of the
activation energies of different diffusion mechanisms reveals the dominant
contribution to self diffusion for different doping and growth conditions
that enter $Q$ via $\Delta\/\mu$ and $\mu_{\mathrm{F}}$. In Table~\ref{tab:Q}
the activation energies for vacancy- and interstitial-mediated self diffusion
are listed.
\begin{table*}
\caption{\label{tab:Q} Activation energy for the self diffusion as mediated by
  vacancies and interstitials (cf. text for details).}
\begin{center}
\begin{minipage}[t]{0.47\linewidth}
\begin{ruledtabular}
\begin{tabular}{lr@{}l@{}lccc}
Defect&\multicolumn{5}{c}{$Q(\mu_{\mathrm{f}},\Delta\/\mu)$ [eV]}&$\mu_{\mathrm{F}}$ [eV]\\
\cline{2-6}
&\multicolumn{3}{c}{explicit}&Si-rich&C-rich&range\\
\hline
V$_{\mathrm{C}}^{2+}$&$6.6$\,eV&$+2\/\mu_{\textrm{F}}$&$-\Delta\/\mu$&6.6--9.0&7.2--9.6&0.0--1.2\\
V$_{\mathrm{C}}^{0}$&$7.3$\,eV&&$-\Delta\/\mu$&7.3&7.9&1.2--$E_{\mathrm{C}}$\\[0.5ex]
C$_{\mathrm{sp}\langle100\rangle}^{2+}$&$6.7$\,eV&$+2\/\mu_{\textrm{F}}$&$+\Delta\/\mu$&6.7--7.8&6.1--7.2&0.0--0.6\\
C$_{\mathrm{sp}\langle100\rangle}^{1+}$&$6.6$\,eV&$+\/\mu_{\textrm{F}}$&$+\Delta\/\mu$&7.2--7.9&6.6--7.3&0.6--0.8\\
C$_{\mathrm{sp}\langle100\rangle}^{0}$&$7.7$\,eV&&$+\Delta\/\mu$&7.2&6.6&0.8--1.8\\
C$_{\mathrm{sp}\langle100\rangle}^{-}$&$9.1$\,eV&$-\mu_{\textrm{F}}$&$+\Delta\/\mu$&7.3--6.7&6.7--6.1&1.8--$E_{\mathrm{C}}$\\[0.5ex]
\hline
Exp.\footnotemark[1] 3C&\multicolumn{5}{c}{$8.72\,\mathrm{eV}\pm0.14\,\mathrm{eV}$}&undoped\\
Exp.\footnotemark[2] 4H&\multicolumn{5}{c}{$7.14\,\mathrm{eV}\pm0.05\,\mathrm{eV}$}&intrinsic\\
Exp.\footnotemark[2] 4H&\multicolumn{5}{c}{$8.20\,\mathrm{eV}\pm0.08\,\mathrm{eV}$}&\emph{n}-type\\
\end{tabular}
\end{ruledtabular}
\end{minipage}
\hspace*{0.04\linewidth}
\begin{minipage}[t]{0.47\linewidth}
\begin{ruledtabular}
\begin{tabular}{lr@{}l@{}lccc}
Defect&\multicolumn{5}{c}{$Q(\mu_{\mathrm{f}},\Delta\/\mu)$ [eV]}&$\mu_{\mathrm{F}}$ [eV]\\
\cline{2-6}
&\multicolumn{3}{c}{explicit}&Si-rich&C-rich&range\\
\hline
V$_{\mathrm{Si}}^{1-}$&$12.2$\,eV&$-\,\mu_{\textrm{F}}$&$+\Delta\/\mu$&11.6--10.4&11.0--9.8&0.6--1.8\\
V$_{\mathrm{Si}}^{2-}$&$13.1\,\mathrm{eV}$&$-2\/\mu_{\textrm{F}}$&$+\Delta\/\mu$&9.6--7.7&9.0--7.1&1.8--$E_{\mathrm{C}}$\\[0.5ex]
Si$_{\mathrm{T\/C}}^{4+}$&$7.8$\,eV&$+4\/\mu_{\textrm{F}}$&$-\Delta\/\mu$&7.8--12.6&8.4--13.2&0.0--1.2\\
 $^{}_{}$                        &$7.7$\,eV&$+4\/\mu_{\textrm{F}}$&$-\Delta\/\mu$&7.7--12.5&8.3--13.1&0.0--1.2\\
Si$_{\mathrm{sp}\langle110\rangle}^{0}$&$9.9$\,eV&&$-\Delta\/\mu$&9.9&10.5&1.1--2.4\\[0.5ex]
\hline
Exp.\footnotemark[1] 3C&\multicolumn{5}{c}{$9.45\,\mathrm{eV}\pm0.05\,\mathrm{eV}$}&undoped\\
Exp.\footnotemark[2] 4H&\multicolumn{5}{c}{$7.22\,\mathrm{eV}\pm0.07\,\mathrm{eV}$}&intrinsic\\
Exp.\footnotemark[2] 4H&\multicolumn{5}{c}{$8.18\,\mathrm{eV}\pm0.10\,\mathrm{eV}$}&\emph{n}-type\\
\end{tabular}
\end{ruledtabular}
\end{minipage}
\end{center}
\footnotetext[1]{Ref.~\onlinecite{hong:79,hon:80}: experiments on poly-crystalline 3C-SiC samples.}
\footnotetext[2]{Ref.~\onlinecite{hong:80,hong:81}: experiments on 4H-SiC single crystals.}
\end{table*}
$Q$ is given as an explicit function of $\mu_{\mathrm{F}}$ and $\Delta\/\mu$.
At Si-rich conditions $\Delta\/\mu$ is 0 and assumes a value of
$-H_{\mathrm{f}}^{\mathrm{SiC}}$ ($-0.61$\,eV in the present calculations) at
C-rich conditions. For the charged defects the range of $\mu_{\mathrm{F}}$
is given in which the particular charge state is stable. The variation of $Q$
with the Fermi-level can be substantial, i.e. for Si$_{\mathrm{T\/C}}^{4+}$ it
amounts up to 5\,eV. For a typical diffusion temperature of 1000\,K the change
of $\Delta\/Q\sim1$\,eV translates into a variation of the diffusion constant
over five orders of magnitude. This striking dependence on the doping
conditions is known as the Fermi-level effect.\cite{tan:92}
Table~\ref{tab:Q} also includes the activation energies deduced from recent self
diffusion experiments for polycrystalline 3C-SiC~\cite{hong:79,hon:80} and for
4H-SiC single crystals.\cite{hong:80,hong:81} To the best of our 
knowledge, these are the first experiments that quantitatively analyzed the self
diffusion constant. A radiotracer diffusion technique was
used to obtain the diffusion profiles.

For the carbon self diffusion our results show a similar importance of the
vacancy-mediated and interstitial-based self diffusion in Si-rich material.
There is no clear dominance of either mechanism. In C-rich material, on the
other hand, the split-interstitial mediated self diffusion dominates for all
doping conditions. Since in compensated and \emph{n}-type material the
split-interstitial and the vacancy are neutral, the Fermi-level effect is
absent. Only in \emph{p}-type material a variation of the activation energy about
1\,eV is present.

The silicon self diffusion in \emph{p}-type and compensated material is mainly
dominated by the silicon interstitials. In \emph{p}-type material the migration is
mediated by the tetrahedral carbon-coordinated interstitial
Si$_{\mathrm{T\/C}}$, both, under Si-rich and C-rich conditions. Here the
metastability of the silicon vacancy suppresses its participation in the self
diffusion. In Si-rich compensated material the silicon spit-interstitial
Si$_{\mathrm{sp\,}\langle110\rangle}$ is relevant.  Only in \emph{n}-type, Si-rich
material a dominant contribution of the silicon vacancy V$_{\mathrm{Si}}$
should be observed. The larger abundance of silicon vacancies under C-rich
conditions leads to its dominance already in compensated material. The
variation of the activation energy of the silicon self diffusion is more
pronounced than for the carbon self diffusion. A Fermi-level effect is
predicted under \emph{p}-type and \emph{n}-type conditions. The variations amount to 2\,eV,
both, in C-rich and Si-rich material.

Experimentally the silicon and carbon self diffusion
was investigated in nominally undoped poly-crystalline 3C-SiC.\cite{hong:79,hon:80} For the carbon
self diffusion an activation energy of 8.72\,eV was obtained. The observed
silicon self diffusion was described~\cite{hon:80} by an activation energy of
9.45\,eV.  From the description of the experiments it is not obvious whether they were carried
out in C-rich or Si-rich conditions. Therefore we compare the experimental
results with our values for C-rich and Si-rich conditions for an intrinsic
Fermi-level. For the carbon self diffusion we obtain an activation energy of
7.3\,eV (V$_{\mathrm{C}}^{0}$, Si-rich) and 6.7\,eV
(C$_{\mathrm{sp}\,\langle110\rangle}$, C-rich), respectively. The experimental
value exceeds our theoretical results by 1.4\,eV--2.0\,eV. This discrepancy
cannot be explained in terms of the missing information about $\Delta\mu$ or a
variation of $\mu_{\mathrm{F}}$. Our results indicate a value of at most 8\,eV.
Only if the interstitial diffusion was completely suppressed during the
experiment and the sample was rather \emph{p}-type than intrinsic, the activation
energy could be explained by that of V$_{\mathrm{C}}^{2+}$. Here the
poly-crystalline nature of the material may also affect the defect equilibrium,
i.e. the grain boundaries may act as sinks for interstitials.  For the silicon
self diffusion we obtain an activation energy of 9.9\,eV
(Si$_{\mathrm{sp}\,\langle110\rangle}$, Si-rich) and 10.4\,eV
(V$_{\mathrm{Si}}^{-}$, C-rich) in intrinsic material. In this case the
experimental value is in good agreement with our results.

Even though our investigation of the defect migration was conducted for 3C-SiC, we
expect that our results may provide a first insight into the diffusion
processes in other polytypes. This expectation is supported by the
finding~\cite{zywietz:99,torpo:01,bockstedte:02} of comparable energetics and
bonding of vacancies in 3C-SiC and 4H-SiC. For example, similar transformation
barriers are found for the neutral silicon vacancy in 3C-SiC (this work) and
4H-SiC.\cite{rauls:00} The main difference is the larger energy gap in 4H-SiC.
Intrinsic conditions, thus, apply in 4H-SiC to $\mu_{\mathrm{F}}\sim1.7$\,eV.
A comparison of our results with the activation energy of the carbon self
diffusion in 4H-SiC single-crystals shows a much better agreement than for the
poly-crystalline 3C-SiC. However, for the silicon self diffusion our
prediction of about 9\,eV substantially deviates from the experimental result
of 7.2\,eV, unless we place the Fermi-level above mid-gap at around 2.3\,eV.
$n$-type 4H-SiC is not covered by our results, since the relevant charge
states of the vacancies and interstitials in 4H-SiC are not present in 3C-SiC.

In the light of our expectation the large variation of the experimental
activation energies between 3C-SiC and 4H-SiC is surprising.  Besides the
poly-crystalline nature of the cubic material, an incomplete point defect
equilibrium or the preferential injection of either vacancies or interstitials
at the interface of the tracer material and SiC may be responsible. Here we
note that the investigation of the aluminum and boron
diffusion~\cite{konstantinov:92,bracht:00} indicate a slower silicon self
diffusion than observed in the tracer diffusion experiments. This finding
inevitably affects the given prefactor and activation energies.

\subsection{Dopant migration}
In SiC, dopant atoms may substitute for carbon or silicon.  For example, aluminum and
phosphorus are known to preferentially bind to the silicon
sublattice~\cite{greulich-weber:97} and nitrogen prefers carbon
sites.\cite{greulich-weber:97} Boron is predicted to bind either to the
carbon or silicon sublattice~\cite{fukumoto:96,bockstedte:01} depending on the
stoichiometry of the sample (C/Si-ratio).  The migration of these dopants is
mediated by mobile intrinsic defects. A mobile vacancy and a substitutional
dopant atom may form a mobile dopant-vacancy-complex. An attractive
interaction within the mobile pair enables migration until the pair eventually
dissociates.  Mobile interstitials may initiate a migration of the dopant by a kick-out reaction.
 
In equilibrium, the dominating mechanism of dopant diffusion depends not only
on the involved migration barriers but also on the abundance of the involved
intrinsic defects. A transient enhanced dopant diffusion that was recently
observed for boron~\cite{laube:99,janson:00} and aluminum~\cite{usov:99} is
triggered by an excess concentration of intrinsic defects, generated for
example by the dopant implantation. Assuming that a similar amount of
interstitials and vacancies is generated, the migration barriers for different
mechanisms are the relevant quantities to compare.  Naturally, for dopants on
the carbon and the silicon sublattice different migration mechanisms apply.
Even though the migration mechanism should certainly be dopant-dependent, some
general conclusions about the dopant diffusion can be drawn from our results
for the diffusion of vacancies and interstitials.

First of all, the availability of the mobile defects depends on the doping
conditions. The discussion in Sec.~\ref{sec:v} showed that a mechanism based
on silicon vacancies is not available in \emph{p}-type material. The metastability of
V$_{\mathrm{Si}}$ in \emph{p}-type material should affect dopant complexes with a
silicon vacancy in a similar way. Thus only the migration of a donor complex
that involves the carbon vacancy should be relevant.  In such a complex the dopant and
the vacancy may be nearest neighbors or second nearest neighbors, when the
dopant substitutes for silicon or carbon, respectively. Whereas the migration
of the second neighbor pair proceeds entirely on the carbon sublattice
by second neighbor hops, the migration of the nearest-neighbor pair
involves an initial exchange of the dopant and the vacancy and further
migration of the vacancy around the dopant.\cite{pankratov:97} This may only
be accomplished by second neighbor hops.
A mechanism based on nearest neighbor hops fails as the necessary
V$_{\mathrm{Si}}$-Si$_{\mathrm{C}}$-complexes are unstable.  In \emph{n}-type
material a mechanism based on silicon vacancies becomes available. With the
same arguments as above a migration will involve only second neighbor
hops of the vacancy.

Consequently, the diffusion of acceptors on the silicon sublattice most likely
is mediated by silicon interstitials in \emph{p}-type material.  More general
conclusions on the role of the mechanisms based on carbon and silicon
interstitials are not possible. Here we only note that carbon-interstitials
and silicon-interstitials (in compensated and \emph{n}-type material) have lower
migration barriers than the corresponding vacancies. The interstitial mediated
dopant migration may be limited by the availability of interstitials and the
barriers associated with the migration of the dopant interstitial.

These arguments are supported by our findings for
boron~\cite{bockstedte:01} and by recent experiments on the diffusion of
boron~\cite{konstantinov:92,laube:99,bracht:00,janson:00} and
aluminum.\cite{konstantinov:92} Our calculations reveal that a migration via
the nearest-neighbor complex B$_{\mathrm{Si}}$-V$_{\mathrm{C}}$ is unlikely.
The barrier of the initial transformation has a similar magnitude as the
transformation of V$_{\mathrm{C}}$-C$_{\mathrm{Si}}$ into V$_{\mathrm{Si}}$ in
\emph{p}-type material.  Furthermore, the nearest-neighbor complex
B$_{\mathrm{Si}}$-V$_{\mathrm{C}}$ is practically unstable in \emph{p}-type material.
The migration of the second-neighbor complex
B$_{\mathrm{C}}$-V$_{\mathrm{C}}$ consists of three hops and involves
migration barriers in \emph{p}-type material of as much as 5.6\,eV, which exceeds the
barrier for the carbon vacancy.  We find much lower migration
barriers for a kick-out mechanism by silicon interstitials (B$_{\mathrm{Si}}$)
or carbon interstitials (B$_{\mathrm{C}}$) and a subsequent migration of the
boron-interstitials.  Though earlier experiments~\cite{mokhov:84} favored a
vacancy related mechanism, the recent quantitative modeling of
boron diffusion experiments~\cite{konstantinov:92,bracht:00,janson:00} attribute the
diffusion to a kick-out mechanism based on silicon interstitials. A similar
conclusion was suggested by co-implantation experiments.\cite{laube:99} Here the transient
enhanced diffusion observed after boron implantation was suppressed by carbon
co-implantation and elevated by silicon co-implantation.

\section{Summary and conclusion}
A picture of the defect migration has been derived from theoretical
investigations based on an \emph{ab initio} method within the framework of DFT.
The analysis of the microscopic structure, the abundance and the migration
mechanisms of interstitials and vacancies contribute to this picture. A strong
influence of the doping conditions is reported for the mobile silicon defects.
Among the silicon interstitials two different interstitial configurations are
relevant, the tetrahedrally carbon-coordinated interstitial in \emph{p}-type material
and the split-interstitial in compensated and \emph{n}-type material. This affects the
interstitial migration in \emph{p}-type material: the migration of the tetrahedrally
carbon-coordinated interstitial is slower than the migration of the
split-interstitial.  A metastability of the silicon vacancy occurs in \emph{p}-type
and compensated material. It transforms into the more stable carbon
vacancy-antisite complex. This transformation of the silicon vacancy is more
likely than its migration. The thermal activation of the reverse process
determines the mobility of the vacancy-antisite complex.  A central result is
the finding that interstitials are more mobile than vacancies.

The importance of interstitials and vacancies in the self and dopant diffusion
is determined by their abundance and diffusivity. The discussion of the two
factors showed that the interstitials play a significant role in the self
diffusion. For instance, silicon interstitials dominate the silicon self
diffusion in \emph{p}-type material.  The dominance of carbon interstitials in the
carbon self diffusion stems from an over-compensation of their relatively low
abundance by their higher mobility.  For the dopant diffusion, similar
qualitative conclusions are drawn and outlined for the case of boron, for
which theoretical calculations indicate an important role of the
interstitial-mediated diffusion.
\begin{acknowledgements}
We acknowledge fruitful discussions with Dr.~G.~Pensl, who also initiated this
project. This work has been supported by the Deutsche Forschungsgemeinschaft,
SFB 292 and SiC Research Group.
\end{acknowledgements}

\end{document}